%% file: main.tex
\pdfoutput=1
\documentclass[sigconf]{acmart}

\copyrightyear{2022}
\acmYear{2022}
\setcopyright{rightsretained}
\acmConference[WWW '23]{Proceedings of the ACM Web Conference 2023}{May 1--5, 2023}{Austin, TX, USA} \acmBooktitle{Proceedings of the ACM Web Conference 2023 (WWW '23), May 1--5, 2023, Austin, TX, USA} \acmDOI{10.1145/3543507.3583311} \acmISBN{978-1-4503-9416-1/23/04}

\input preamble/preamble
\input preamble/macros

\ccsdesc[500]{Information systems~Online advertising}
\ccsdesc[500]{Information systems~Traffic analysis}
\ccsdesc[300]{Networks~Online social networks}

\keywords{User Web Tracking, Facebook Click ID, Web Pixels}

\begin{document}

\title{The Hitchhiker's Guide to Facebook Web Tracking with Invisible Pixels and Click IDs}

\author{Paschalis Bekos}
\affiliation{%
 \institution{University of Crete/FORTH}
 \country{Greece}
}

\author{Panagiotis Papadopoulos}
\affiliation{%
\institution{FORTH}
\country{Greece}
}

\author{Evangelos P. Markatos}
\affiliation{%
\institution{University of Crete/FORTH}
\country{Greece}
}

\author{Nicolas Kourtellis}
\affiliation{%
\institution{Telefonica Research}
\country{Spain}
}

\renewcommand{\shortauthors}{Paschalis Bekos, Panagiotis Papadopoulos, Evangelos P. Markatos, \& Nicolas Kourtellis}

\input{sections/0_abstract}

\maketitle

\input{sections/1_introduction}
\input{sections/2_background}
\input{sections/3_pixel_adoption}
\input{sections/4_fbtracking_mechanics}
\input{sections/5_fbtracking_experiment}

\input{sections/6_IDs_3rdparties}

\input{sections/7_expiration_dates}
\input{sections/10_related_work}
\input{sections/11_discussion}

\section*{Acknowledgements}
This project received funding from the EU H2020 Research \& Innovation programme under grant agreements No 830927 (Concordia), 830929 (CyberSec4Europe), 871370 (Pimcity), 871793 (Accordion), 101021808 (Spatial), and 883543 (CC-DRIVER).
These  results reflect only the authors' view and the Commission is not responsible for any use that may be made of the information it contains.

\bibliographystyle{unsrt} 
\balance
\bibliography{main}
\appendix
\newpage
\input{sections/appendix_extended_results}

\end{document}

%% file: preamble/preamble.tex
\usepackage{booktabs}  
\usepackage{pbox}
\usepackage{epsfig,endnotes}
\usepackage{epstopdf}
\usepackage{xcolor}
\usepackage{graphicx}
\usepackage{arydshln}
\usepackage{xspace}
\usepackage{url}
\usepackage{breakurl}
\usepackage{comment}
\usepackage{balance}
\usepackage[utf8]{inputenc}
\usepackage{graphicx}
\usepackage{amsmath}
\graphicspath{ {./figs} }
\usepackage[tableposition=top,font={small}, skip=7pt]{caption} 
\usepackage{eurosym}
\usepackage{listings}
\usepackage{multirow}
\usepackage{float}
\usepackage{csquotes}
\newcommand{\ts}{\textsuperscript}
\usepackage{float}
\usepackage{tabularx}
\usepackage{adjustbox}
\usepackage{multicol}
\usepackage{lipsum}
\usepackage{subcaption}

\usepackage{color}
\definecolor{lightgray}{rgb}{.9,.9,.9}
\definecolor{darkgray}{rgb}{.4,.4,.4}
\definecolor{purple}{rgb}{0.65, 0.12, 0.82}

\lstdefinelanguage{JavaScript}{
  keywords={typeof, new, true, false, catch, function, return, null, catch, switch, var, if, in, while, do, else, case, break},
  keywordstyle=\color{purple}\bfseries,
  ndkeywords={class, export, boolean, throw, implements, import, this},
  ndkeywordstyle=\color{darkgray}\bfseries,
  identifierstyle=\color{black},
  sensitive=false,
  comment=[l]{//},
  morecomment=[s]{/*}{*/},
  commentstyle=\color{purple}\ttfamily,
  stringstyle=\color{red}\ttfamily,
  morestring=[b]',
  morestring=[b]"
}

%% file: preamble/macros.tex
\newcommand{\point}[1]{\vspace{0.01in}\par\vspace{0.01in}\noindent{\textbf{#1:} }}

\newcommand{\FBCLID}{FBCLID\xspace}
\newcommand{\FBCLIDs}{FBCLIDs\xspace}
\newcommand{\eg}{{e.g.,}\xspace}
\newcommand{\etc}{{etc.}\xspace}
\newcommand{\etal}{{et al.}\xspace}
\newcommand{\ie}{{i.e.,}\xspace}

\newcommand{\dataSize}{17K\xspace}

\newcommand{\FB}{FB\xspace}

\newcommand{\FBpixel}{\FB Pixel\xspace}
\newcommand{\FBP}{\FB Pixel\xspace}
\newcommand{\FBC}{\_fbc\xspace}

\newcommand{\fbp}{\_fbp\xspace}
\newcommand{\fbc}{\_fbc\xspace}

\newcommand{\TrancoFBP}{2.3K FBP\xspace}
\newcommand{\SampledFBP}{990 FBP\xspace}
\newcommand{\TopTranco}{T10\xspace}
\newcommand{\SampledTranco}{6S\xspace}

%% file: sections/0_abstract.tex
\begin{abstract}

Over the past years, advertisement companies have used various tracking methods to persistently track users across the web.
Such tracking methods usually include first and third-party cookies, cookie synchronization, as well as a variety of fingerprinting mechanisms. 
Facebook (\FB) (now Meta) recently introduced a new \textit{tagging} mechanism that attaches a one-time tag as a URL parameter (namely \FBCLID) on outgoing links to other websites. 
Although such a tag does not seem to have enough information to persistently track users, we demonstrate that despite its ephemeral nature, when combined with \FBpixel, it can aid in persistently monitoring user browsing behavior across i) different websites, ii) different actions on each website, iii) time, \ie both in the past as well as in the future.
We refer to this online monitoring of users as \FB web tracking.

We find that \FBpixel tracks a wide range of user activities on websites with alarming detail, especially on websites classified as sensitive categories under GDPR.
Also, we show how the \FBCLID tag can be used to match, and thus de-anonymize, activities of online users performed in the distant past (even before those users had a \FB account) tracked by \FBpixel.
In fact, by combining this tag with cookies that have rolling expiration dates, \FB can also \textit{keep track of} users' browsing activities in the future as well.
Our experimental results suggest that 23\% of the 10k most popular websites have adopted this technology, and can contribute to this \textit{activity tracking} on the web.
Furthermore, our longitudinal study shows that this type of user activity tracking can go as far back as 2015. ).
Simply said, if a user creates for the first time a \FB account today, \FB could, under some conditions, match their anonymously collected past web browsing activity to their newly created \FB profile, from as far back as 2015 and continue tracking their activity in the future.

\end{abstract}

%% file: sections/1_introduction.tex
\section{Introduction}
\label{sec:intro}

Undoubtedly, advertising is the primary source of revenue for Facebook (\FB, now Meta)\cite{bid-prices}.
Recent reports suggest that Meta had 117.9 billion revenue in 2021~\cite{facebook-statistics},
97.5\% of which was generated from advertising of third parties on \FB and Instagram~\cite{third-party-adv-revenue}.
This revenue was achieved due to millions of ad campaigns whose success relied, in part, on key behavioral data and \FB user tracking.

Indeed, \FB uses cookies produced by its \FBpixel for user activity tracking purposes.
This \emph{third-party}, JavaScript-based \FB code sets a \emph{first-party} cookie (\textbf{\fbp}) on a website and transmits it back to \FB to track the visitor.
However, it cannot actually point to the \emph{real} identity of the user: it is a cookie unique per browser and website, but it is not the \emph{real} identity of the user\footnote{\emph{Real identity}: The user identity as is known to \FB: \ie the \FB account of the user.}.
Potentially to amend this issue, \FB recently (2018) introduced the \FB Click ID (\FBCLID): a \emph{one-time} tag that \FB appends to outbound links (\eg ads, websites, \etc) that are shared on its platform.
When a user clicks on an ad (or other external article) on \FB, the outbound request includes a \FBCLID query parameter that is transmitted to the landing page of the third party.
This parameter is a long string that looks random and changes each time a user clicks on an outbound link. Even if the same user clicks twice on exactly the same link (after page reload), the two \FBCLID\/ values generated are different from each other\footnote{Example of actual link with \FBCLID: \url{https://www.ncbi.nlm.nih.gov/pmc/articles/PMC5678212/?fbclid=IwAR0J2ueFwGP2ZSIznw04PQEFAbkMDue3T9YSg6}}.
Since the \FBCLID seems to have a temporary one-time value, one might assume that \emph{it cannot be used to persistently track users} across the web and time. 
On the contrary, in this work, we show that \FBCLID, despite its ephemeral nature, can aid \FB in de-anonymizing users, whose activities on websites (that support \FBpixel) have been persistently tracked.
In fact, as we show later, \FBCLID\/ can be used in conjunction with \FBP to reveal the \emph{real} identity of a user and attribute all the previously captured browsing activity of an unknown \fbp, to a real \FB account. 

With this work, we are the first to shed light on the \fbp and \FBCLID tagging mechanisms, assess their functionalities and how they collaborate, and the implications on users’ privacy and persistent activity tracking on the web from \FB.
To understand the extent of this tracking by \FBpixel and the opportunity of \FB to de-anonymize users' activity with \FBCLID, we analyze a total of \dataSize websites of a wide range of categories and ranking in the top 1M websites, and the website behavior when the \FBCLID tag and \fbp are present. 
We showcase the importance of the \FBCLID tag which could eventually reveal the real identity of users and link them to past anonymous activities on various categories of websites.
In fact, we find significant presence of \FBpixel on sensitive category websites under GDPR.
In addition, our longitudinal study shows that this type of behavioral tracking on websites with \FBP dates as back as 2015, or even as 2013 (when the precursor of this technology was first introduced by \FB).
This means that \FB could eventually have built customer profiles for its users with browsing activity, on websites utilizing \FBpixel, captured as far back as 2013.
We show that those tags and cookies presented as first-party cookies can be used to persistently track users with rolling expiration dates, providing an adequate replacement of the third-party cookies blocked by privacy browsers or plugins.

%% file: sections/2_background.tex
\section{Background}
\label{sec:background}

\subsection{What is the Facebook Pixel?}

Tracking users who visit a website has been traditionally done with third-party cookies.
Thus, \FB Pixel (\FBP)~\cite{FacebookPixel} was first introduced in 2013 as an analytics tool for \FB to help advertisers (websites external to \FB) measure and increase the effectiveness of their advertising campaigns.
\FB also uses \FBP to track users' browsing activity anonymously while outside the platform.
Given recent regulations, restrictions and blockade of third-party cookies, advertisers rely on other mechanisms such as \emph{first-party} cookies to continue tracking user activities and provide better insights on ad campaigns.
Those cookies are placed by third-party JavaScript code on the website~\cite{outdated-JS-libraries,third-party-js-to-first-party-website}, and allow tracking the activity of a user under a pseudonym but not their real identity.
This is a major drawback on tracking as it renders the attribution of history of activities to a real user (not a pseudonym) almost impossible.
\FB replaced the \FBpixel mechanism by a second version in 2015 and further updated it in Oct. 2018 to better leverage first-party cookies that \FBpixel can inject per website.
Then, \FB introduced \FBCLID to enable \FB\/ to track, not just browsers, but \FB user accounts.

\FBP is a piece of JavaScript code added to a website 
as a graphic element with dimension $1\times 1$ pixel that is loaded when a user lands on the website hosting it. When \FBP is included in a website, it also has embedded a URL pointing to \FB servers, with a specific ID reflecting the specific \FBP account holder, and can be  used by \FB to track relevant audiences for brand ads~(\cite{replug}).
When a user visits a \FBP-enabled website, an instance of the pixel loads in the HTML code of the page on the browser. According to \FBP documentation~\cite{Pixel-fbp-cookie}, 
if a corresponding first-party \textit{\_fbp} cookie does not exist on the browser, one is created and a unique ID is saved for this domain.
\FBP is responsible for reporting to \FB the user activity for the duration of the visit.

The value of \fbp cookie has the following format:
\begin{equation*}
\textcolor{blue}{version.subdomainIndex.creationTime.RandomNumber}
\end{equation*}
\vspace{-1mm}
Example of the \_fbp cookie:
{\footnotesize
\textcolor{blue}{\texttt{fb.1.1596403881668.1116446470}}
},
where:
\begin{itemize}
    \item \textit{version:} always the prefix \textit{fb}
    \item \textit{subdomainIndex:} domain where the cookie was defined. \eg `com':0, `shoes.com':1, `www.shoes.com':2
    \item \textit{creationTime:} UNIX time in milliseconds when the \_fbp cookie was created
    \item \textit{RandomNumber:} a number generated from \FBpixel’s SDK ensures that every \_fbp cookie is unique.
    As stated by \FB, it is ``generated by the Meta Pixel SDK to ensure every \_fbp cookie is unique''~\cite{Pixel-fbp-cookie}.
\end{itemize}

\begin{figure}[t]
    \vspace{-2mm}
    \includegraphics[width=.86\columnwidth]{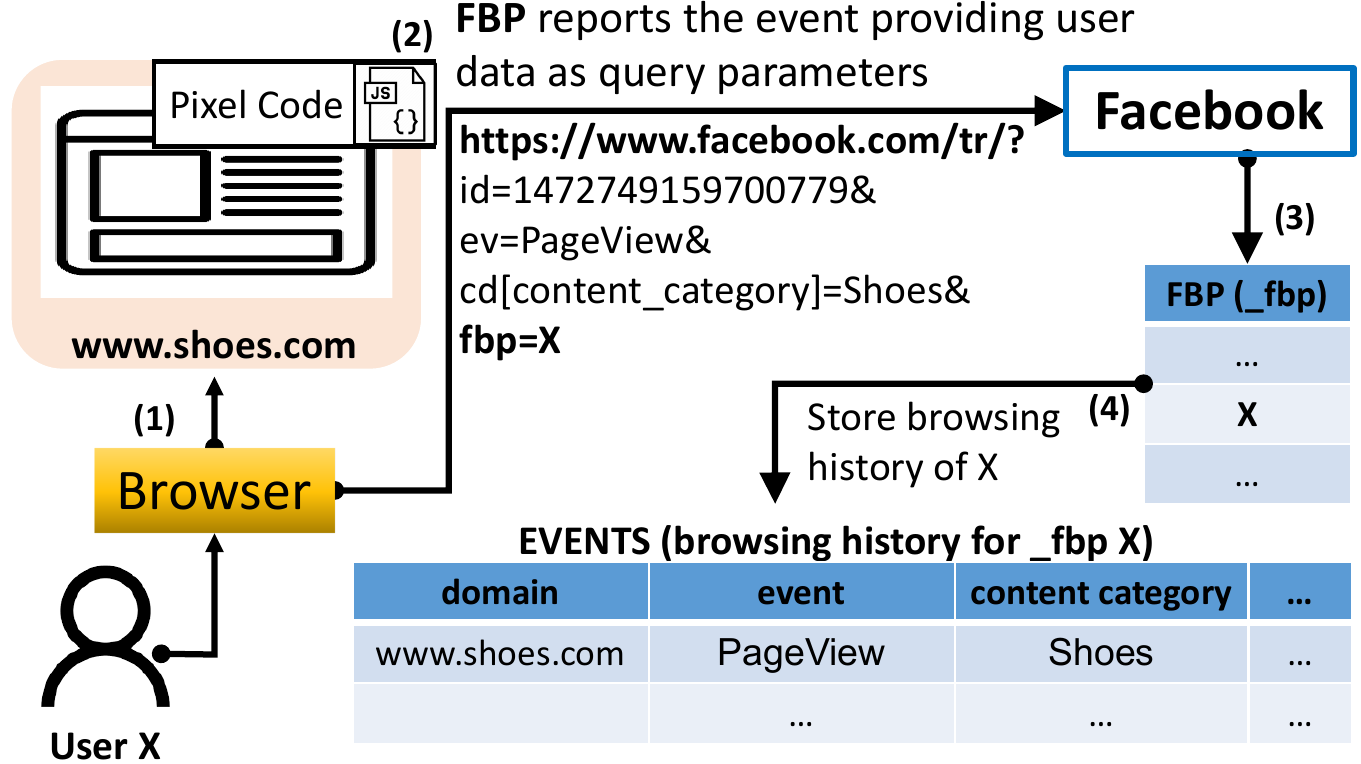}
    \vspace{-2mm}
    \caption{How \FB can record the browsing activity of user X into an anonymous \FB profile using reporting from \FBpixel.}
    \label{fig:FBP-to-FB-for-matching}
    \vspace{-3.5mm}
    
    \Description{
        Figure 1 describes the process of recording browsing activities for a user yet unknown to Facebook.
        Assume this unknown user X visits a website namely www.shoes.com which utilizes Facebook Pixel (FBPixel). Facebook Pixel, stores a unique id for this user as the first party cookie _fbp. The visitation event (Page-View) also triggers Facebook's Pixel reporting process which sends to Facebook the event and the _fbp cookie for this unknown yet to Facebook user. 
        Based on the uniqueness of _fbp, Facebook can store the information that received under _fbp and keep collecting the future browsing activity of X under this anonymous id.
    }
\end{figure}

\subsection{How does \FBP track users?}

\FBP performs \textit{event-driven} tracking of user activities on websites.
That is, the webpage administrator has to define to \FBP the behavior they wish to track and report to \FB, by defining \textit{events}, \ie actions, that a user takes on the website.
Events can be general such as `Page-View', or specific such as `Time-On-Page'.
Actions the website has configured to be tracked are defined as \textit{conversions}.
\FB provides a set of 18 trackable events named \textbf{Standard Events} (listed in Appendix ~\ref{sec:appendix-events}).
\FB is informed about actions being tracked by \FBpixel and triggered by the user, allowing \FB to build a \textit{history of activities} for said user.

The \textit{Base Pixel code} is a small segment of JavaScript code which acts as an `initiator' for the overall \FBP behavior.
The base function \textit{fbq()} is initialized and a library of functions currently called \textit{fbevents.js}~\cite{current-version-pixel-library} is loaded on the website (the older version was called \textit{fbd.js}~\cite{previous-version-pixel-library}; both versions have been also found in~\cite{third-party-scripts-in-the-web}).
This library is the core mechanism of the \FBP.

In Figure~\ref{fig:FBP-to-FB-for-matching}, we illustrate the reporting process of the Page-View (visitation of the website) event.
When an event is fired on a web-page because a user X loaded the website (step 1), \FBP reports to \FB information about the event and user that caused it, which initially is the \fbp value (for an unknown user, step 2).
On the other hand, \FB can record the reported information in appropriate tables (steps 3 and 4) for future use in ad-conversion and further user profiling and tracking, thus building a history of activities for a (yet) unknown user on this website.
Based on the uniqueness of the \fbp cookie, the browsing activities tracked by \FBP will be attributed to this (yet) unknown \fbp value.

Developers of the website can also declare \textbf{Custom Events} (apart from the 18 standard events) they wish to track, as well as additional user information they wish to share with \FB through \FBpixel (see Appendix \ref{sec:appendix-events}, Table~\ref{tab:user-data-fb} for the additional ids). 
However, the baseline information that is received is the event (\eg PageView in Figure~\ref{fig:FBP-to-FB-for-matching}) and some IDs that help \FB keep a history of activities and potentially match this history to a \FB user, as shown later.

%% file: sections/3_pixel_adoption.tex
\begin{table}[t]
\caption{\FBpixel adoption in websites of sensitive categories.}
\vspace{-2.5mm}
\footnotesize
\begin{tabular}{lccccc}
\toprule
               & health & politics & sexual orientation & ethnicity & religion\\
               \midrule
Pixel adoption & 16.9\% & 14\% & 10.5\% & 5.3\% & 4.6\% \\
Total websites & 7679   & 5663 & 1140 & 2509 & 9119 \\
\bottomrule
\end{tabular}
\label{tab:sensitive-categories}
\Description{
    Pixel adoption across 5 GDPR sensitive categories (health, politics, sexual orientation, ethnicity, religion). Adoption rate calculated with respect to the size of each category list of websites, showcasing that Facebook Pixel is more adopted in Health and Politics than Sexual Orientation, Ethnicity and Religion websites.}
\vspace{-3.5mm}
\end{table}

\section{\FBpixel Presence on the Web}
\label{sec:adoption}

First, we measure the extent to which \FBP has been adopted across the web, before analyzing further its impact to users' privacy.
We also quantify \FB's ability to build elaborate histories of user activities on sites, by tracking a variety of events users perform.

\subsection{\FBP Adoption in Top 1M}
\label{sec:top10K-regular-crawl}

\subsubsection{Experimental Setup}

To find the adoption of \FBP, we crawl top websites and analyzed any \FBpixel-related cookies,
along with the network traffic produced.
For this crawling, we use the \textit{Tranco} List~\cite{tranco-ranking} of top 1M websites, ranked for 2.11.21 (ID:L394)~\cite{cite-to-download-this-list}.
To make experiments computationally feasible, we use the top 10K websites~(\TopTranco), as well as a sample of 1000 websites for each of the following 6 rank ranges: 10K-20K, 20K-50K, 50K-100K, 100K-200K, 200K-500K, 500K-1M, for a total of 6K extra websites~(\SampledTranco).
For this crawling, we follow state-of-the-art practices and use Chromium browser~\cite{papadogiannakis2022googleids, papadogiannakis2021postcookie, chen2021swap-cookies, sanchezrola2021cookieecosystem} and Puppeteer~\cite{Puppeteer} on several parallel VM instances.
Moreover, given that automated crawling can face network errors, disconnections, unavailability of web servers, \etc, we visit each website three times.
Each visit is performed with a clean browser instance.
In order to emulate a real user more closely, we do not perform headless crawling.
Instead, we launch a 1600x1200 pixel instance of Chromium. All crawlings performed, followed the ethical guidelines described in Appendix \ref{sec:appendix-ethics}.

Finally, we consider the navigation to each page completed by monitoring two main indicators from Puppeteer:
1) \textit{networkidle0:} No more than 0 network connections for at least 500 ms;
2) \textit{domcontentloaded:} The DOMContentLoaded event is fired (in order not to wait for style-sheets, images, \etc, to load, that would slow down our crawling process).
If both indicators are true,  we assume that the navigation for the specific website has been completed.
After that, we check all cookies stored from this domain for any whose name matches the pattern \textit{\_fbp}.
If such a cookie exists, we store the whole array of cookies for that website. 
Each  visit to a new website is done with a new instance of browser in order to avoid cross-domain contamination.
This crawl was completed on 17.01.22.

\subsubsection{Top websites with \FBP in 2022}

Our results indicate that 2,308~(\TrancoFBP) of the top 10K websites (23.08\%) used the \FBP. 
It is interesting to note that this percentage (23.08\%) is in line with previously reported results: Chen \etal (see~\cite{chen2021swap-cookies}, Table 3) found 23.77\% of the top 10K Alexa websites store \fbp\/ cookies.
We also look into the \SampledTranco set,
and find that 990 (or 16.5\%) of websites adopt \FBP(\SampledFBP), indicating that higher-ranking websites are more eager to adopt \FBP.
This is not surprising since such websites are expected to be more active in ads and, therefore, in collaborations with \FB.
We conclude that 1 in 5-6 top websites employ \FBP and its first-party cookie, enabling \FB to track users' activities on websites, across a large portion of the Web.

\begin{figure}[t]
    \includegraphics[width=.85\columnwidth]{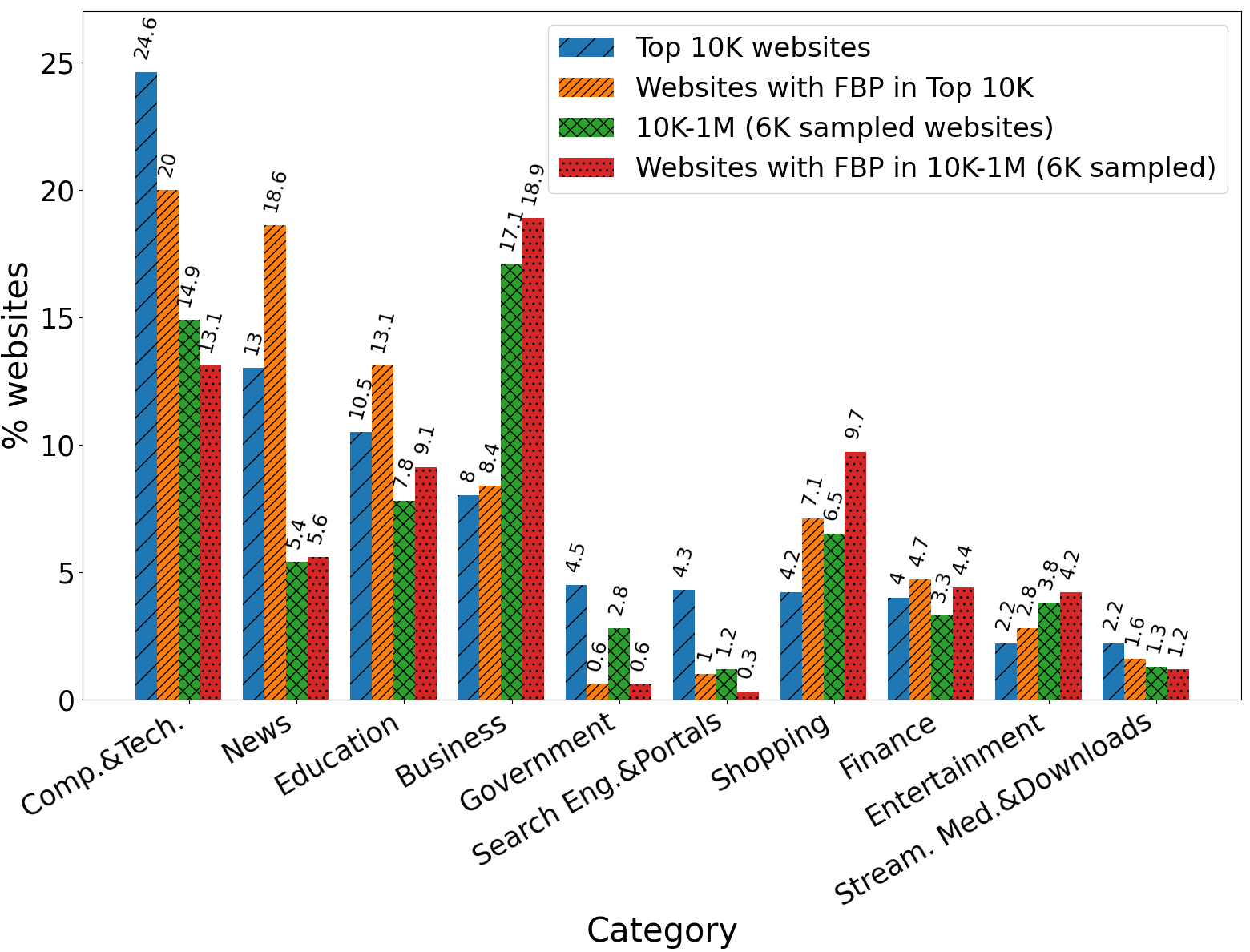}\vspace{-.2cm}
    \caption{Top 10 categories depicting the proportional rate of \FBP adoption in \TopTranco and \SampledTranco sets.}\vspace{-.4cm}
    \label{fig:GeneralWebVSPixelCategories}
    \Description{
        Adoption rate of Facebook Pixel across the most popular categories for the two sets Top 10K and 6K sampled websites collected from the Tranco list.
        With this figure we answer the following question: If an online user visits a website that utilizes Pixel across the top 10K (or 6K sampled) what is the most common category that they will land on? 
    }
\end{figure}

\subsubsection{Website categories with \FBP}

Next, we quantify what types of websites are more eager to adopt \FBP and its associated tracking.
In order to classify the websites found to operate \FBP, we use Cyren's URL category check gate~\cite{cyren-URL-category-check-gate}.
Figure~\ref{fig:GeneralWebVSPixelCategories} shows the 10 most popular categories across websites for \TopTranco and \SampledTranco, as well as for websites adopting \FBP for each set (\TrancoFBP and \SampledFBP, respectively).
The results suggest that 20\% of the \FBP-adopting websites are categorized as ``Computers and Technology" and 18.6\% of them are ``News" websites.
This should not be surprising as these kinds of websites are very frequent on the web and offer lots of opportunities for ad sales. Figure~\ref{fig:GeneralWebVSPixelCategories} also shows whether websites of a specific category are more aggressively adopting \FBP than the average site.
Indeed, websites in ``News", ``Education", and ``Shopping" are disproportionately more eager to adopt \FBP.
In contrast, and perhaps expected, websites under ``Government'' and ``Search Engines \& Portals'' are less eager to adopt \FBP, as they may be less dependent on ads.
We also observe some deviations from the lower ranking websites, which employ \FBP more than expected in additional categories of ``Business'', ``Finance'', and ``Entertainment''.
Alternatively, if we examine how many websites in a given category are adopting \FBP, the top 3 categories are ``Shopping'', ``News'' and ``Education'' (more details in Appendix~\ref{sec:appendix-categories-adoption}).

\subsubsection{Pixel adoption in sensitive categories}
\label{sec:sensitive-cats}
Next, we focus on specific categories deemed sensitive under GDPR~\cite{identifying-sensitive-urls-curlie}, and study the adoption of \FBpixel in each.
For this, we crawl Curlie's Internet directory~\cite{curlie} and fetch various websites for the following 5 \textit{sensitive} categories: ethnicity, health, politics, religion, sexual orientation, thus constructing 5 lists with unique domains for each category.
We visit each website utilizing the same method as earlier to identify the use of FB Pixel (we focus on this script compared to other FB-related scripts studied in~\cite{who-tracks-sensitive-domains}).
Table~\ref{tab:sensitive-categories} shows per sensitive category the size of each list, and portion of websites using \FBpixel.
Interestingly, websites on health, politics and sexual orientation are quite eager to adopt \FBpixel and track users' activities.

\subsubsection{Tracking with Standard \& Custom Events}
\FB provides 18 standard events that trigger the \FBpixel reporting process.
Each website can also define its own events to be reported through \FBpixel, by modifying the call to the \textit{fbq()} base function.
The tracking of an event is achieved by modifying the 1\ts{st} argument of the function with possible modifications: `\textit{track}', `\textit{trackCustom}', `\textit{trackSingle}', `\textit{trackSingleCustom}' and the order of the next arguments, as described by \FB~\cite{advanced-tracking,custom-conversions}.
To understand the intensity of user tracking via different events across websites, we perform automatic static analysis of the pages that use \FBpixel, and identify the aforementioned function and the tracked events included therein.
This methodology allows us to go beyond recent work which collected \FBpixel events by observing the traffic of real users but on a handful of websites(~\cite{markup-pixel-hunting}), and instead scale the crawling to thousands of websites.
In fact, and inspired by recent work on under-study of subpages~\cite{counting-internal-pages}, we go beyond the homepage of each website and study its subpages as well:
For each homepage, we extract all links pointing to the same domain, and keep all that are one level below the top-level domain.
Then, on 27/09/2022 we crawl all identified homepages and its subpages to detect (statically) the events tracked.

\begin{figure}[t]
    \centering
    \includegraphics[width=\columnwidth]{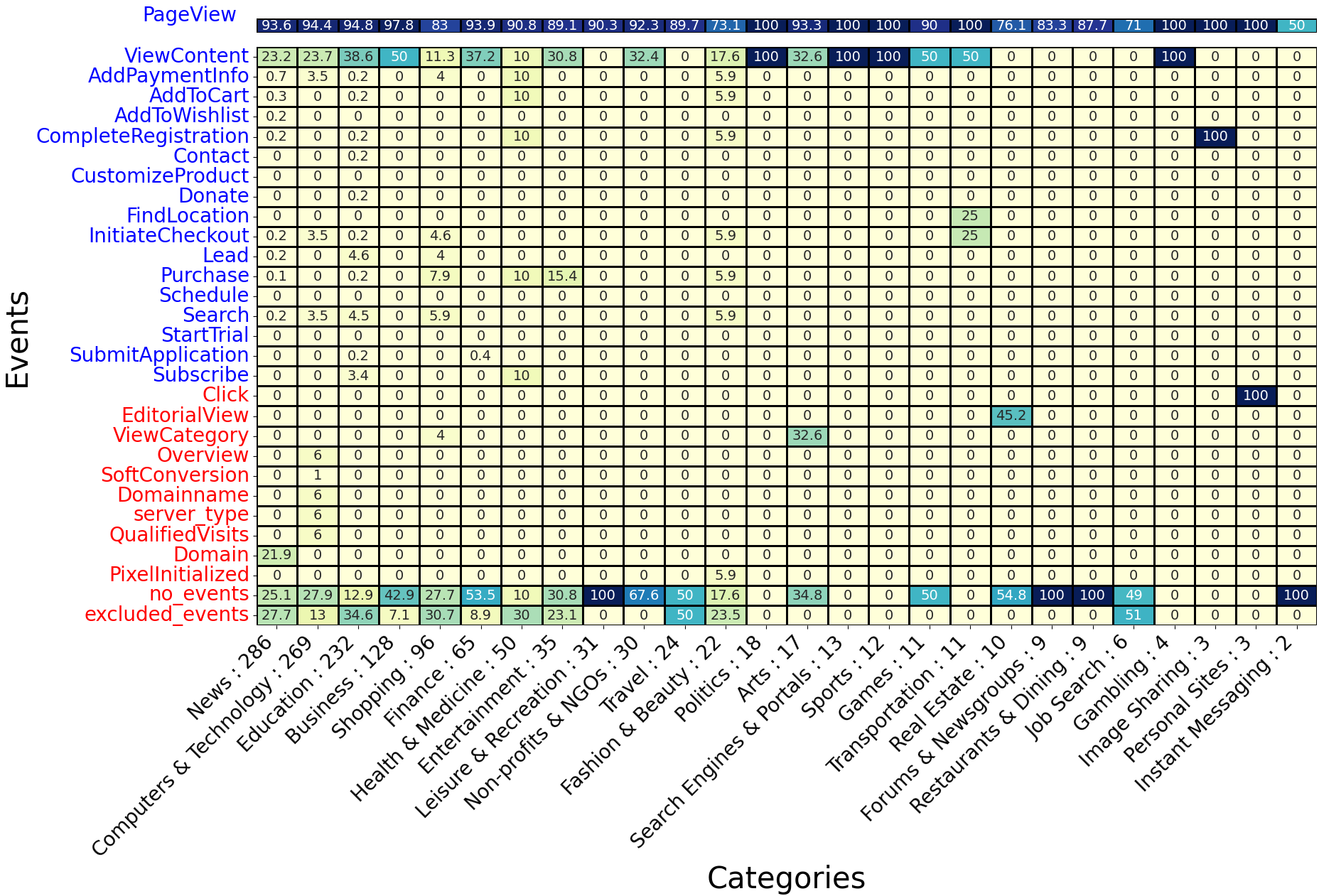}\vspace{-.2cm}
    \caption{Events tracked by \FBpixel on subpages of top 10K websites, analyzed per category. Blue-colored events are the 18 standard from \FB. Red-colored events are newly discovered and studied. Cells show portion of websites per category (x-axis) tracking said event (y-axis).}
    \label{fig:events-per-category}
    \Description{
        Heatmap depicting the portion of Standard and Custom events captured in sub-pages of the websites utilizing Facebook Pixel across the Tranco list's top 10K. Portions are normalized with respect to websites' categories.}\vspace{-.45cm}
\end{figure}

From the \TrancoFBP websites,
we find 1473 that load \FBP statically on their HTML.
In total, we crawled 1473 homepages and 18683 subpages (out of which only 932 did not track any event).
We extract the fbq() calls and identify the events set to be tracked by \FBP.
From our analysis, beyond the 18 \FB standard events, we discover 10 new events, excluding website-specific events encountered only in one website, or variables.
From the standard events, we did not encounter: \textit{CustomizeProduct}, \textit{StartTrial} and \textit{Schedule}.
We normalize the events captured in the websites per category.
Figure ~\ref{fig:events-per-category} shows the portion of events (per category) found in subpages (due to space limits, results on homepages in Appendix~\ref{sec:appendix-events}, Figure ~\ref{fig:events-per-category-homepages}).

Unsurprisingly, in both homepages and subpages, the great majority of websites track the \textit{PageView} event.
Beyond that, there is a wide variety of events tracked and reported by different categories of websites.
For example, `Fashion \& Beauty' and `Shopping' websites, beyond the actual content viewed by their visitor, also track and report to \FB financial transactions on their site related to payment, shopping cart, etc.
Other interesting outliers are `Personal sites' that track the `click' event, `Imagesharing' websites that track `completeregistration', `Education' websites that track `lead', `search', `subscribe', and `Non-profits\&NGOs' websites that track `addtocart', `lead', `purchase', `completeregistration'.
Alarmingly, and in addition to Sec.~\ref{sec:sensitive-cats} findings, `Health and Medicine' websites track events such as `add payment info', `addtocart', `purchase', `completeregistration', and `subscribe', all reported to \FB.
Overall, the variety of events reported to \FB by its pixel, enables it to build a detailed history of activity per user, and finally profile the users more effectively for better ad-targeting.

\subsection{Historical \FBP Adoption}
\label{sec:historical-adoption}

Next, we want to know when websites started adopting \FBP.
For this, we proceed as follows:
We visit all top 5K (from Tranco) websites using the \textit{Wayback Machine} (WM)~\cite{waybackmachine} to find if they had \FBP code embedded in their HTML, and when this was done for the first time, starting from 01.2022 and going back to 2013.

\subsubsection{A decade of tracking through \FBpixel}

\begin{figure}[t]
    \centering
    \includegraphics[width=0.9\columnwidth]{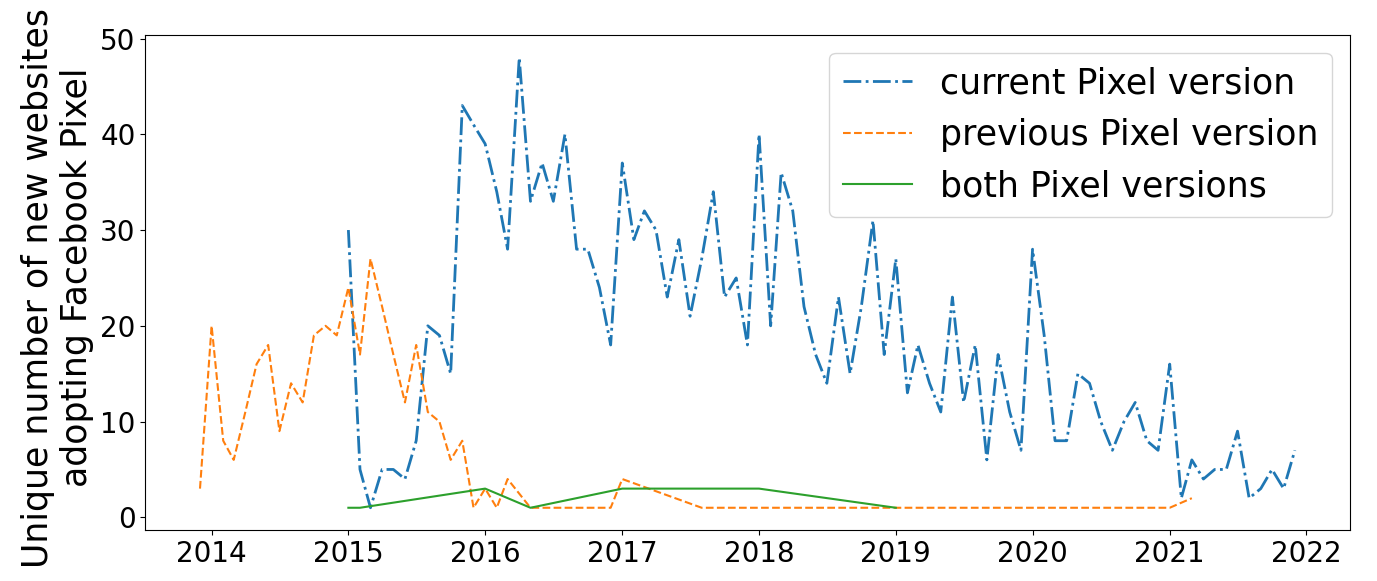}\vspace{-.1cm}
    \caption{New websites with \FBP, through time, for the top 5K websites.
     We see that the previous Pixel version was adopted around 2014 while the current Pixel version was introduced around 2015.
     }
    \label{fig:UniquePixelAdoptionPerMonth-top5K}
    \Description{
        Historical adoption rate of both versions of Facebook Pixel (uniquely and combined). Rates depict quarterly spikes that seem to conform with marketing campaigns of Facebook or website updates.
    }
    \vspace{-5mm}
\end{figure}

Figure~\ref{fig:UniquePixelAdoptionPerMonth-top5K} shows the distribution of unique new websites adopting \FBP through time, for the top 5K websites (similar results were found for the general top 1M websites and are omitted due to space).
First, we note that \FBP has been around since 12.2013 with the 1\ts{st} version, while its adoption (per new website) peaked around early 2015.
At the same time, in 2015, we also had the release of the 2\ts{nd} version, whose adoption peaked in early 2016, while the 1\ts{st} version lost momentum.
In fact, \FB faced-out the 1\ts{st} version in 02.2017. Second, we notice a few websites using both versions in parallel, as evident from the third line.
This is probably due to websites that adopted the code for the new \FBP version without removing the previous version. Forth, as noted by the spikes per month, \FBP adoption was not smooth, but in waves due to \FB marketing campaigns and website updates, happening at the beginning of each year and quarter.

\subsection{Key Takeaways}
About 23\% of \TopTranco websites facilitate \FB tracking activities, using \FBP for their ad conversion analytics.
We find \FBP in 40+ categories of websites, with the top 3 being Computers \& Technology, News, and Education.
This wide coverage shows that conversion tracking through \FBP can capture activity of different profiles of users.
We also find at least 28 different web events tracked and reported by \FBP (10 more than the Standard events of \FB).
Of special interest are GDPR-declared \textit{sensitive} categories such as health, politics and sexual orientation that track user activity via various events.
\FB has been tracking users' browsing activities since at least 2013 using \FBP on websites, building elaborate profiles with a wide range of captured actions.
Thus, the opportunity for tracking and de-anonymizing a user's browsing history of activities for a website that utilizes \FBP goes back a decade or so.

%% file: sections/4_fbtracking_mechanics.tex
\section{\FBpixel Tracking Mechanics}
\label{sec:fbtrackingmechanics}

Next, we focus on the key question: \textit{How can \FB use its recently launched one-time-tag \FBCLID with its \fbp cookie to match historical website visitors with \FB user profiles?}
In the following paragraphs, we explore the mechanics of these cases.

\subsection{Matching website visitor coming from \FB}
\label{sec:case1}

This is the base scenario, which involves a user with a \FB profile, who arrives at website W after they clicked on a post or ad of W while browsing within \FB.
In most such cases, there is a \textit{fbclid} (\FBCLID) query parameter with an ID appended in the URL of the landing website.
If W runs \FBP, the pixel saves the appended \FBCLID under a cookie named \textit{\_fbc}, with the following format:
\begin{equation*}
\textcolor{blue}{version.subdomainIndex.creationTime.fbclid}
\end{equation*}
\noindent where the fields are similar to the \fbp cookie apart from the last, which holds the \FBCLID value.
An example of such a \FBCLID is:
\begin{center}
\textcolor{blue}{\footnotesize{\texttt{IwAR0X0CXJ2hZB\_CNGJbBmjHYZqZiS9-MJ9QrFxJBRkacHhJMMjnQLsEjvXYZ}}}
\end{center}
Since \FB knows who performed the click, we assume that \FB assigns the \FBCLID to the user.
Thus, with a browsing history of the user for website W under the unique \fbp value, the combination of the two cookies (\ie\/ \_fbp and \_fbc\/) enables \FB to uniquely track \FB users and their visits through time, inside or outside \FB.

\point{\FBCLID generation}
Interestingly, although the \FBP has been in use in its latest version since 2015, \FBCLID has been appended to outgoing URLs only since mid-October 2018, with no official documentation from \FB~\cite{fbclid-webpage}.
Since \FB does not reveal how these hashed-like strings are generated or what information is stored inside them, we provide some empirical analysis on their usage and possible values.
A \FBCLID parameter has 61 alphanumeric characters which can consist of upper/lower case letters, numbers and symbols (\eg '-'); thus, it is not base64 encoding.
\FB for Developers documentation also states~\cite{Pixel-fbp-cookie} that regardless of the existence of said cookies, if the \FBCLID parameter exists in a  URL, the ID can be sent to \FB to provide browsing insights for the \FB user.

\point{\FBCLID assignment}
Given the lack of documentation on how these IDs are being generated and assigned to users, ads, links, \etc, from within \FB, we conduct a few experiments to acquire better understanding.
We perform these experiments using two different user accounts on \FB, and three different browsers (Google Chrome, Mozilla Firefox and Microsoft Edge).
We inspect the HTML code of \FB's home page (for those accounts) and in the \FB pages of some businesses.
We observe that \FBCLID is assigned inside the \textit{href} value of HTML elements, meaning these values are not created ``on the fly'', but rather they are assigned once the document is loaded.

Upon further examination of the \FB webpage JavaScript source code, we find that when loading a page inside \FB, there is a \textit{static} array called \textit{click\_ids}, loaded as part of a JSON object inside a <script> tag, whose values are updated upon every request that is landing on the specific page (\eg reload).
This array hosts 50 \FBCLID values, each with a fixed length of 61 characters.
Therefore, these \FBCLIDs cannot be uniquely distributed amongst the different URLs that could exceed 50 in total.
Also, there can be the same \FBCLID assigned to different URLs.
But, the same \FBCLID seems to only be assigned to elements of same class name, thus concluding \FBCLID is probably element dependent.

\begin{figure}[t]
    \centering
    \includegraphics[width=\columnwidth]{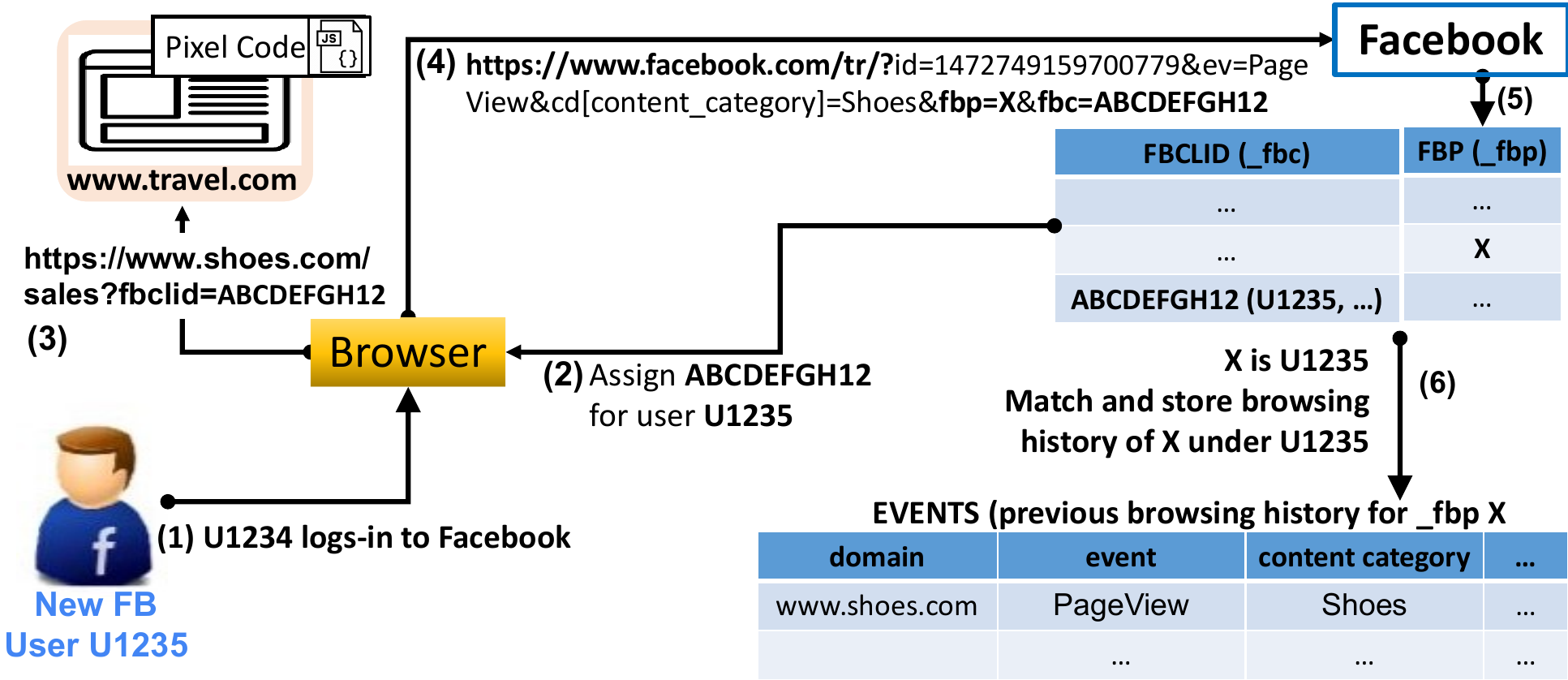}\vspace{-0.1cm}
    \caption{
    How \FB can match the browsing of the anonymous profile of user Z, using \FBCLIDs, with the existing \FB user U1234.}
    \label{fig:FBP-to-FB-for-matching-fbclids}
    \vspace{-4mm}
    \Description{
        Ability for attribution of browsing history of activities for an unknown user (_fbp) to a Facebook profile through Facebook Click ID (FBCLID) tags. The FBCLID is assigned to the user within the platform to a Facebook user and reported back from a website through Pixel, along with the _fbp cookie, which could point to the previously recorded history of activities.
    }
\end{figure}

\point{\FB User to Visitor Matching}
Let us assume that a user starts with a clean browser (no cookies) and visits the website \textit{www.travel.com}.
Also assume this website uses \FBP, i.e., \FB\/ can collect some browsing history information about this user, but under a pseudonym \textbf{Z} (not the real identity) corresponding to the \fbp value.

Let us now assume that this user actually has a \FB\/ account and is known to \FB\/ as user \textbf{U1234}.
Let us also assume that this user will now login into \FB\/ - into their account.
Figure~\ref{fig:FBP-to-FB-for-matching-fbclids} shows what happens from this  point on and how \FB\/ could  match the browsing history of \textit{Z}, inside travel.com, with the \emph{real identity} of \FB\/ \textit{U1234}.
\textit{U1234} browses through their newsfeed on \FB\/ (step 1) which contains an external link to website \textit{www.travel.com}, possibly as the result of ad-targeting.
When \textit{U1234} clicks on this URL
(step 3), a \FBCLID value of \textbf{ABCDEFGH12} is also appended on the outgoing URL (step 3).
This value (\ie\/ \textit{ABCDEFGH12}) was set by \FB\/ on step 2, and then \FBpixel on \textit{www.travel.com} creates a \fbc cookie storing the \FBCLID locally.
This visit also triggers an event such as the ``PageView'' on the webpage  that \FBP picks up.
Then, \FBP performs a GET request to a \FB server providing the URL information about this event as a query parameter (step 4).
The event that triggered this request is appended in the URL, along with both the \fbp and \fbc cookie values, all reported to \FB (step 4).
Given that the reported \FBCLID \textit{ABCDEFGH12} was given to \textit{U1234}, and given that this value 
(\ie  \textit{ABCDEFGH12}) arrived to \FB\/ along with the \fbp cookie for \textit{Z}, \FB\/ can now infer that user \textit{Z} and user \textit{U1234} are the same person.
As a result, all the previously performed browsing activity of \textit{Z} captured on www.travel.com, can now be attributed to user \textit{U1234}.

\subsection{Matching website visitor with new \FB user}
\label{sec:case3}
Given that \FBpixel reports the browsing activity of a user regardless of the existence of a \FB account, the same build-up of history of activities can be achieved for non \FB users.
If such a user later on creates a \FB account and browses websites with \FBpixel, then the same matching process can occur and connect the previous browsing history of the new \FB user to a specific website W.
To depict the overall matching process, we envision a 3-day scenario:

\noindent \textbf{1st day:} \textit{Z} visits a website (say \textbf{www.travel.com}) with \FBpixel, which informs \FB\/ of the visit. \FB\/ may store information regarding the user behavior on the website, under anonymous user \textit{Z}.

\noindent \textbf{2nd day:} The user creates a \FB\/ account. From now on the user is known to \FB as \textbf{U1234}. At this point, \FB\/ still does not have enough information to associate \textit{U1234} with the browsing history of \textit{Z}.

\noindent \textbf{3rd day:} While \textit{U1234} browses their \FB newsfeed (step (1), Figure ~\ref{fig:FBP-to-FB-for-matching-fbclids}), they click on a link pointing to \textit{www.travel.com}.
This link (step 3) takes as argument the \FBCLID value \textbf{ABCDEFGH12}. The visit also triggers a ``PageView'' event on the page that \FBP reports to \FB, thus enabling the aforementioned the matching process.

%% file: sections/5_fbtracking_experiment.tex
\section{\FBpixel User Active Profiling}
\label{sec:steps-to-match-users}

Next, we quantify \textit{how many \FBP-enabled websites, in conjunction with \fbp and \FBCLID, can enable \FB to reconstruct a user's browsing history, even before the user had created a \FB account?}

\subsection{Experimental Setup}
We generalize the process outlined in Section~\ref{sec:case3} into 4 steps, to scale the data collection process on multiple websites:

\point{Step S1} Similar crawling as in Sec.~\ref{sec:adoption}, but on \TrancoFBP websites.
\point{Step S2} Similar to S1, but with 2 differences: (1) 
we load only \fbp, and no other to avoid loading session cookies; (2) while waiting for navigation to complete, we monitor all GET requests to \FB that include \fbp as a query parameter with its value, allowing \FB to build a browsing history for each site, under a user ``pseudonym''.

\point{Step S3} Extracting a real \FBCLID value from a dummy \FB account.
We store this ID and use it in the next crawling, in S4.
\point{Step S4} Similar to S2, but with a fundamental difference.
Now we visit each of the \TrancoFBP websites with the \FBCLID (extracted in S3) appended in the URL of each website.
Each visit is done by loading the \fbp cookie corresponding to each website and monitoring network traffic, as in S2.
Because of the \FBCLID, a \fbc is created and sent to \FB along with \fbp cookie.
We capture the outgoing traffic that includes both \fbp and \fbc values. 
We store each URL with those values for further analysis.

\subsection{Browsing history build-up \& matching}

Looking into the traffic collected from S2, we find that out of 2,308 websites storing the \fbp cookie when visited (S1 and S2), 2,223 (or 96\%) websites report this event to \FB.
Thus, \FB can create an anonymous profile for user Z with these websites included, along with any other useful data such as type of website, \etc

At the end of S4, \FB has both the \fbp (unique browser ID) and \fbc containing the one-time \FBCLID tag.
Thus, by looking into the outgoing traffic of S4, we can find on how many websites \FB can link to Z's newly created \FB profile.
In fact, from the 2,308 websites, 2,165 (or 93\%) send to \FB both the original \fbp and the newly created \fbc values.
In particular, 92.3\% of the websites report both \fbp (S2) and \fbc (S4), meaning that for those websites the previous browsing activity was tracked based on the \fbp value and could be matched to a user based on the \FBCLID value of S4.
The remaining portion is due to cases where websites do not track the visitation event, or do not load \FBpixel due to network errors, or use URL shorteners that cut \FBCLID.

\subsection{Key Takeaways}

The above results show how the great majority of top websites report user-related IDs facilitating persistent tracking of anonymous users by \FB, even \textit{before} \FB user profile creation.
In fact, given this tracking on websites by \FBP is event driven, it only takes one click of the user on a given website, for \FB to be able to attribute the user's browsing history to an (existing) \FB user: 96\% of \FBP-enabled websites report ``PageView'' (visitation) event to \FB and 93\% of such websites report both \fbp and \fbc values, enabling \FB to perform user matching with anonymous historical data.

%% file: sections/6_IDs_3rdparties.tex
\section{\FBP: is Anyone Listening?}
\label{sec:fbp-fbclickid-third-parties}

Next, we look to answer the following question: 
\textit{Do websites share \FB-related IDs with other third parties, thus, facilitating or enhancing user behavioral tracking beyond the scope of \FB?}

\subsection{Shared \fbp with third parties}

From S1 and S2 crawls of Sec.~\ref{sec:steps-to-match-users}, we find 4 websites that report \fbp cookies to other third parties, beyond \FB.
Interestingly, 3 are sub-domains of the first party, but appear to be advertising or analytics APIs.
Thus, we cannot know if they propagate these IDs to other vendors after receiving them.
Repeating this analysis on the \SampledFBP sites,
we find 13 more third parties receiving \fbp cookies.

These results confirm our intuition that other third parties can perform the same level of profiling as \FB, by recording the user browsing history through time and attempting to link anonymous user browsing with legit user profiles using external IDs.
These results also confirm past work~\cite{chen2021swap-cookies, sanchezrola2021cookieecosystem} that similarly found third parties receiving \fbp cookies.
For example,~\cite{chen2021swap-cookies} found 76 third parties receiving the \fbp value.
The core difference with our results is that~\cite{chen2021swap-cookies} used dynamic taint analysis to capture data flows of third-party JavaScript at runtime, whereas we capture GET requests that include those cookie values in the request URL.

\subsection{Shared \FBCLID with third parties}

Next, we measure the network activity that a \FBCLID value yields when visiting a website, to study how websites handle this ID.
In order to confirm if this functionality is affected depending on the \FBCLID value, we craft 3 versions of this ID, as follows :\textit{real} (a value extracted from a real \FB user account), \textit{random} (a \FBCLID-looking value, following the proper format, but with randomly picked characters), \textit{dummy} (a value that does not follow proper format, with string value: ``Adummy\_param'').
These variations allow us to study if a website's \FBP checks the value of \FBCLID passed by \FB, and performs some filtering, or if it blindly records it in the \FBC cookie, and then alerts \FB of the user's arrival.
Furthermore, we study if the website performs any actions such as forwarding to other third parties, depending on the type of \FBCLID.

We revisit each of the \TrancoFBP (and \SampledFBP) websites, capturing every GET request encapsulating either the \FBCLID value injected on the URL, or the \fbp cookie value recorded for each website.
For each of the 3 versions of the ID, the crawler:
1) Visits the website with the appended \FBCLID parameter and value on the URL.
2) \textit{first\_hop:} Gets all URLs from the HTML DOM that include this value and all URLs from the network that receive those values with a GET request (if any).
These outgoing URLs, apart from \FB properties, can be third parties informed of these \FB IDs.
3) \textit{second\_hop:} Visits all such detected URLs and captures again all outgoing requests that include the \FBCLID or \fbp values.
Clearly, this is an artificial invocation of said URLs, but we want to measure if these third parties pass the received IDs to further third parties.

We find the total number of third parties receiving the \FBCLID value, per case is: 1931 for real, 1920 for random, and 1920 for dummy \FBCLID.
Thus, websites do not check \FBCLID values: they pass IDs to third parties indiscriminately of its value, its structure, \etc, as long as it is passed into a URL parameter named ``fbclid''.

\begin{figure}[t]
    \centering
    \includegraphics[width=0.95\columnwidth]{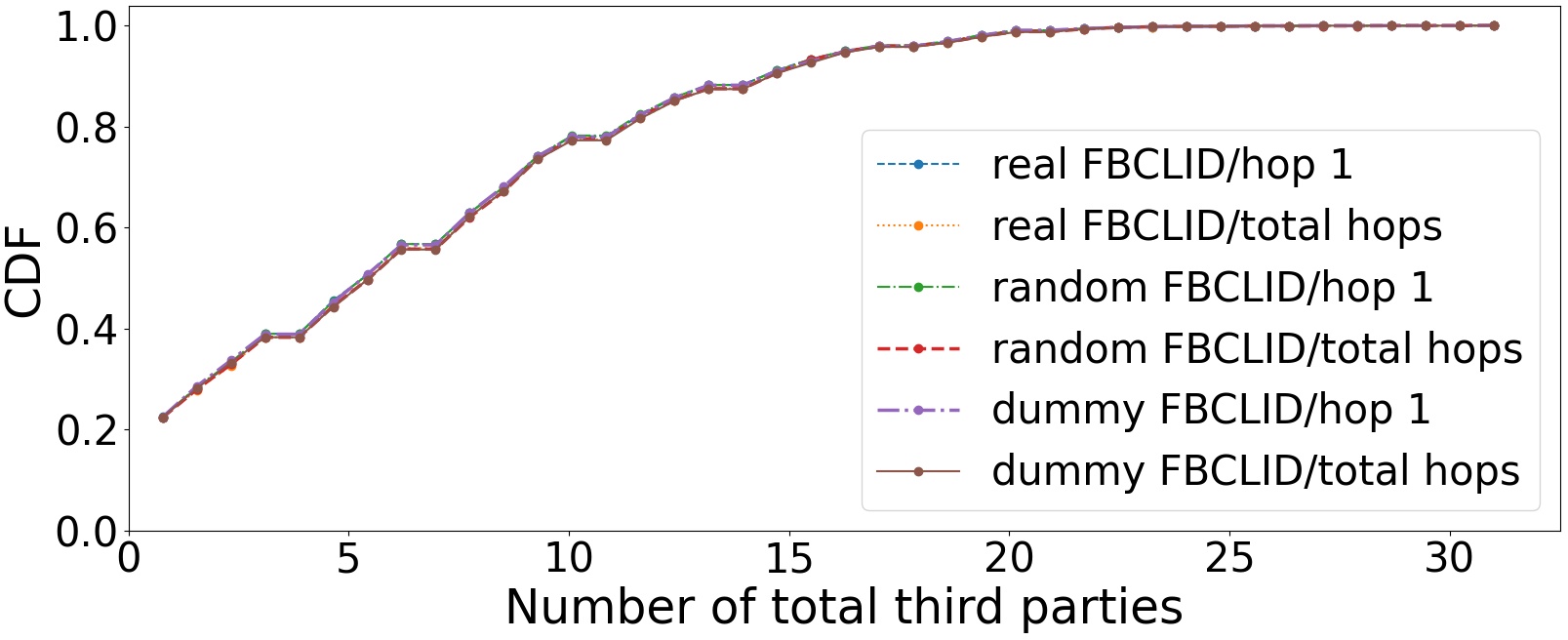}\vspace{-0.8mm}
    \caption{CDF of number of unique third-parties informed of \FB-related IDs, per hop within \TrancoFBP websites.}
    \label{fig:FbclidCDFs}
    \Description{
        Cumulative distribution function (CDF) of unique third parties per hop, receiving each Facebook Click ID (FBCLID) value (real, random, or dummy) for the websites utilizing Pixel in the top 10K of the Tranco list. By unique third party, we refer to a third party domain that was not encountered in an earlier hop. The plot indicates that neither FBPixel nor websites handle the FBCLID tag differently under variations of the value.
    }
    \vspace{-3.5mm}
\end{figure}

\begin{figure}[t]
    \includegraphics[width=0.9\columnwidth]{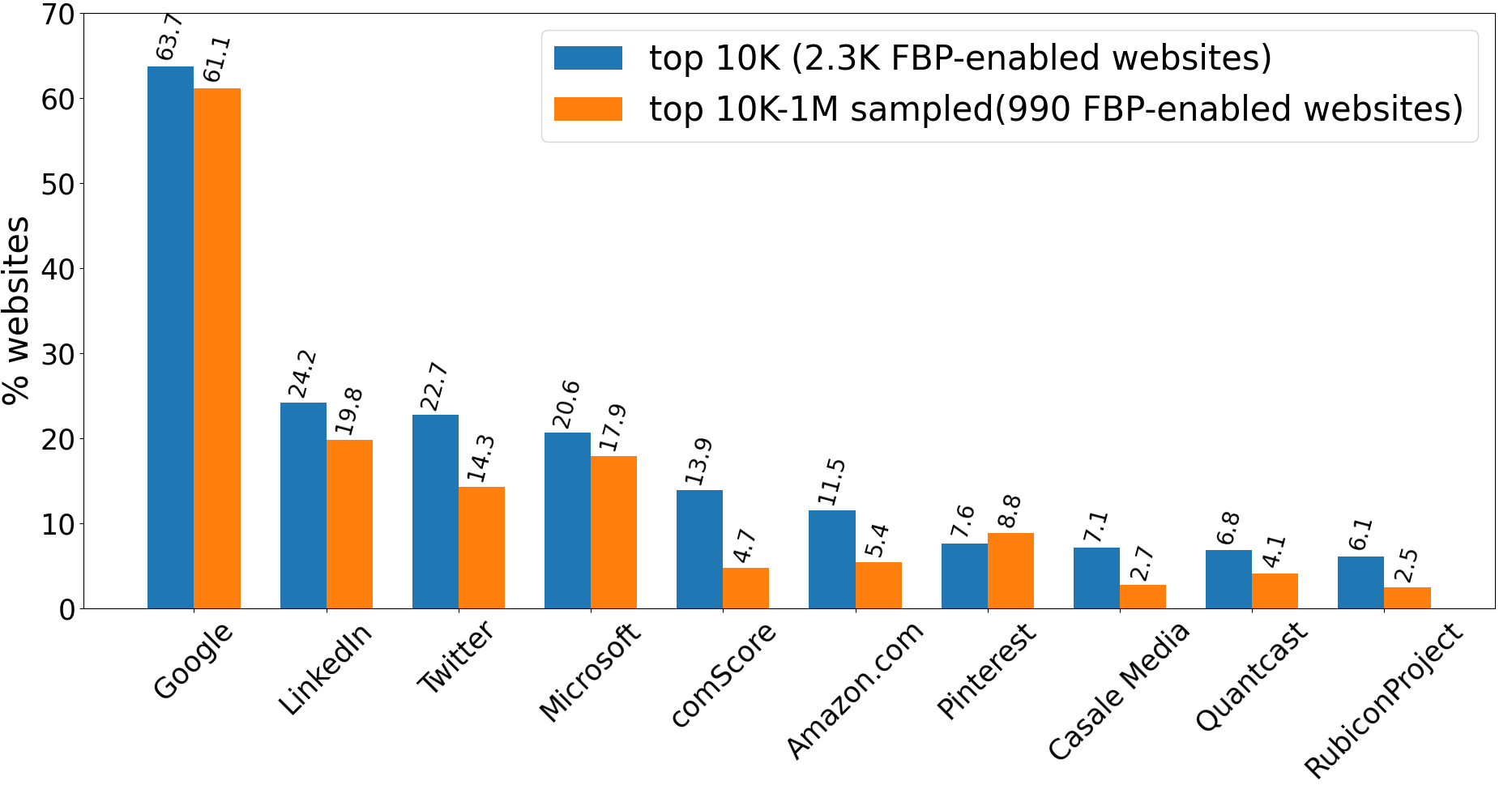}\vspace{-.3cm}\caption{
    Top 10 third-party entities from each list of \FBpixel-enabled websites (\TrancoFBP and \SampledFBP).}
    \vspace{-.5cm}
    \label{fig:top-third-parties-entities}
    \Description{
        Top 10 third party entities receiving the FBCLID values for the websites utilizing Pixel across the top 10K and the sampled 6K websites utilizing Pixel lists. Most predominant entities are Google, LinkedIn and Twitter.
    }
\end{figure}

\point{Unique vs. Total third parties}
Figure~\ref{fig:FbclidCDFs} shows for the 3 cases, the distribution of unique 
third parties for `first hop' and `total hops' (as explained earlier), excluding \FB-related domains, in the \TrancoFBP websites (results for the \SampledFBP can be found in Appendix~\ref{sec:appendix-FBCLIDsharing-third-party-distribution}, Figures ~\ref{fig:FbclidCDFSampled} and~\ref{fig:FbclidDefaultDistributionPlotSampled}).
We find that 22.4\% (24.4\%) of websites from \TrancoFBP list (\SampledFBP) do not send \FBCLIDs to any third party.
However, a median website passes these IDs to 6.2 (4.9) third parties.
In the extreme scenario, there are websites that pass \FBCLIDs to a maximum of 31 (26) third parties.
Thus, lower ranked websites seem to share \FBCLIDs with fewer entities than popular websites, at the median and maximum case.

\subsection{Top Third Parties and Entities}

Having the traffic of each \FBCLID value for the two lists (\TrancoFBP and \SampledFBP), we merge every third party encountered from each list, excluding every subdomain of the websites visited (\eg metrics.shoes.com is excluded as a first-party subdomain of shoes.com), ending with 1398 third-party domains.
Focusing on their Top Level Domain (TLD) using the tld Library~\cite{python-tld},
we find 755 unique TLDs, many of which are typical such as doubleclick.net (Appendix~\ref{sec:appendix-FBCLIDsharing-third-party-distribution}, Fig.~\ref{fig:top-third-parties} plots the top 10 third parties receiving \FBCLID).

Next, we use the \texttt{Disconnect.me} list 
~\cite{disconnect-me} to match these domains with third-party entities.
We match 210 out of 755 (27.8\%) third-party domains to various services of those entities, and Figure~\ref{fig:top-third-parties-entities} shows the top 10 of them.
Google is the top, with more than 61\% of websites reporting to its services user \FBCLIDs.
Interestingly, Microsoft can be considered second, since it also owns LinkedIn, with Twitter and ComScore following.
Finally, the breakdown of third parties and their entities is very similar between the two sets of websites (\TrancoFBP vs \SampledFBP).
This means that such trackers and other Web entities well embedded across all ranks of websites get informed of \FB IDs and thus \FB user activity.

\subsection{Key Takeaways}
\FBpixel does not have a mechanism to distinguish real vs. artificial values of \FBCLIDs: if there is a URL parameter \textit{fbclid}, its value is reported to \FB.
Also, a median website passes such IDs to 4.9-6.2 third parties beyond \FB, who can persistently record the user browsing activities using the \fbp, and attempt to link anonymous user browsing with legit user profiles using external IDs, thus allowing a more elaborate analysis of a user behavior.

%% file: sections/7_expiration_dates.tex
\section{Rolling Expiration Dates of \FBpixel}
\label{sec:expiration-dates}

Next we investigate: \textit{How long does this behavioral tracking by \FB last on websites?}
Intuitively it should be for the 3-months lifespan of the \FBP cookies (\fbc and \fbp)~\cite{facebook-cookie-policy}.
But is it true?

\subsection{Experimental Setup}
To answer this question, we measure how often \fbp cookies update their expiration date, and thus, how long \FB can track the same user under the same (anonymous or not) profile.
Thus, we study if this persistent ID (\fbp) stays alive beyond 90 days, by:
\point{Step S1} Crawling as Sec.~\ref{sec:adoption} (navigation settings, \etc).
We store \fbp cookie and record its expiration date.
\point{Step S2} Reloading all \TrancoFBP websites to fetch \fbp again.
\point{Step S3} Reloading all such websites, but with the \FBCLID appended in each URL, forcing each website to generate a \FBC cookie.
We store the produced \fbp cookie, but not the \FBC cookie since it is not persistent (\ie the value changes per \FBCLID).

This experiment aims to establish when websites update the expiration date of their \fbp, depending on the actions performed on the website: reload vs. reload with a \FBCLID included, \ie\ the visit originated from within \FB.
As a sanity check, in all steps and websites that a cookie is created and stored, we confirm that the expiration date is 90 days after the creation date.
\subsection{Rolling tracking: cookies that never die}
The great majority of \TrancoFBP (1942 or 84.1\%) update the \fbp expiration date on every event (reload or with \FBCLID).
Also, there are 115 (or 5.0\%) websites that do not update the expiration date of \fbp, regardless of event or action on the website.
Also, $\sim$3.2\% outlying cases have the expiration date of the \fbp cookie updated upon \textit{only} one specific event.
While \FBpixel handles these cookies, they still remain first party cookies, thus websites have access to modify them.
Thus, a small deviation between their behavior on different websites can be expected.

Interestingly, there are 172 (7.5\%) websites that blocked our access after a specific event, or stored duplicate results for an event.
Not generating a \fbp value could be due to network errors (unable to load \FBpixel), or the specific website stopped supporting \FBpixel since our first crawl in Sec.~\ref{sec:adoption}.
Given that for this experiment we did not use any mechanism to disguise our crawler from bot detection, duplicate cookies could be due to network errors, redirection to human authentication mechanisms (\eg CAPTCHA), or even blocking the crawler from further action if it was flagged as a potential bot.

\subsection{Key Takeaways}

Overall, 87.3\% of \TrancoFBP websites update the expiration date of \fbp cookie, under some visiting condition (\eg reloading website, visiting with \FBCLID, \etc).
This continuous updating of \fbp expiration enables \FB to perform persistent and consistent tracking of users across visits on the same website, browser and device, beyond the initial 90-day expiration date of \fbp cookies.

%% file: sections/10_related_work.tex
\section{Related Work}
\label{sec:related-work}

User tracking on websites by first and third parties, primarily done for ad-conversion attribution and user profiling purposes, has been extensively studied in the last decade~\cite{mayer2012third, nikiforakis2013cookieless, acar2014web, 10.1145/3355369.3355582, englehardt2016online, falahrastegar2016tracking, le2017towards, papadopoulos2017if, papadopoulos2018cost, papadopoulos2019cookie, solomos2020gdpr-changes, chen2021swap-cookies, papadopoulos2017long, agarwal2020stop,  sanchezrola2021cookieecosystem,demir2022first-party-cookie-field,information-flows-ads}.
Research also detected changes in user web tracking due to enforcement of EU regulation such GDPR and e-Privacy on EU-based websites, by measuring changes in cookies and other fingerprinting technologies employed by websites and third parties~\cite{sanchezrola2019gdprcookie, utz2019informed, solomos2020gdpr-changes, matte2020cookie-banners-choice, papadogiannakis2021postcookie,tracking-block-study}.
In such studies, \FB is identified as a key player in the third-party Web tracking ecosystem.
Previous \FB-related works~\cite{the-like-button-1,the-like-button-2,third-party-detecting-defending,shadow-profiles} focused on how \FB track a user's online activities across the Web through social widgets (the `like' button) and third-party cookies.

Studies on trackers' historical presence on the web~\cite{Libert2015ExposingTH,Wambach2016TheEO,wayback-study} looked at the prevalence of various third-party tracking techniques over a period of time up to 2016.
As a complement to those studies, we showcase the expansion of Analytical Tracking (\FBpixel) up to 2022.
Recent studies on cookies that are injected by third parties (such as \FB) into a browser, but stored as first-party cookies~\cite{chen2021swap-cookies, sanchezrola2021cookieecosystem,  demir2022first-party-cookie-field, dambra2022tracking-from-users-perspective} identified the entities and their relationships in the current post-third-party ecosystem.

Moreover, studies~\cite{fouad2018trackingpixels, online-tracking-pixel-ref, dambra2022tracking-from-users-perspective} focused on the analysis of such pixels and identifying such tracking techniques and their prevalence in the web.
The Markup~\cite{markup-pixel-hunting} recently conducted the first large-scale study to measure the presence of \FBpixel and the data it collects from real users.
They analyze \FBpixel-reported IDs and how \FB can extract personal information from the hashed values.
Also~\cite{markup-pixel-hunting-findings-1,markup-pixel-hunting-findings-2} raised issues regarding \FB tracking of 4M college students applying for federal financial aid, by sharing personal data with \FB through \FBpixel.
Lastly,~\cite{health-advertising-on-facebook} studies how cancer-related health companies use third-party tools to track patients' behavior between their websites and \FB, capturing the \FBCLID as data shared with \FB.

\point{Our contributions}
While previous studies captured a broad view of the first-party tracking ecosystem (and key participation of \FB in it), they do not explore how de-anonymization can be achieved for those anonymous first-party trackers.
Moreover, they do not explore how mechanisms such as \FBpixel could replace methods like social widgets, that leverage third-party cookies, effectively achieving user de-anonymazation.
We study how this pair of cookies can aid \FB to persistently build an elaborate history of user activities, and match anonymous web users (via their activities outside \FB) to \FB users.
In particular, we are the first to study the following questions:
(a) How \FB can keep track of a variety of activities for a user that has no \FB account yet?
(b) How \FB can match previous browsing history and activity on a given website (utilizing \FBpixel) to a new or existing \FB user?
(c) How long can this behavioral tracking of an online (\FB or not) user last?
(d) How can the \FBCLID parameter help \FB, through \FBpixel, and its persistent cookie \fbp, to match a user's specific online activity on a website (\eg\ viewing, clicking, purchasing, \etc) to a certain \FB profile, and even possibly expose \FB activity of a user to other third parties?

%% file: sections/11_discussion.tex
\section{Conclusion}
\label{sec:discussion}

\FB\/ has recently introduced \FB Click ID (\FBCLID): a one-time tag passed as argument to all outgoing URLs for users who browse through \FB\/.
Although this one-time tag is ephemeral and seemingly cannot be used for tracking, with our present study, we show for the first time how \FBCLID can be combined with \FBpixel, a conversion tracking tool embedded via JavaScript on websites, and be used to track users' activities both in space (\ie\/ in websites utilizing \FBpixel), and in time (\ie\/ in the past and future).
Although \FBpixel advertises a 3-month-long lifespan and can seemingly limit any tracking to at most 3 months, unfortunately, in a great majority of websites with \FBpixel, the pixel employs rolling expiration dates for the first-party cookies it places (\fbp), which can postpone its lifespan (and its associated tracking) indefinitely.
We have experimentally verified this behavior on more than 20\% of top 10K, and 6K samples of top 10K-1M websites.
We also find that \FBpixel has been tracking user activity since 2013 and is present in many sensitive under GDPR categories of websites.
Finally, these \FB-related IDs are passed to several third parties beyond \FB, aiding other entities to monitor and profile users' web activity.
\point{Disclaimer} This work reports our findings on \textit{what} data are been collected and how they \textit{could be} used.
Unfortunately, we do not know how, when and why they are used after been collected.

\point{Countermeasures}
Regardless of privacy regulations (GDPR, e-Privacy), and efforts to limit third-party tracking (\eg\ Chrome and Safari browsers blocking cross-session cookies), companies heavily relying on ads come up with novel ideas to preserve their ad reach and profits.
For example, as shown here, with tracking through the \FB Pixel utilizing first-party cookies and URL tags that evade third-party cookie blockade.
Privacy-focused browsers have been a subject of research~\cite{tracking-free-browser}, and recently Brave and Firefox~\cite{brave,firefox} adopted URL striping to avoid URL tagging such as with \textit{fbclid} parameters.
Still, websites can share other IDs such as hashed emails, etc., thus, enabling \FB and other third parties of historical matching.

Ad-blockers which block requests fetching third-party libraries (in our case the \FBpixel core library) into a site could be a solution.
But this can also be bypassed by redirecting the request that loads the library through proxy servers or similar techniques.
Indeed, as previous studies showed (e.g.,~\cite{evading-adblockers-2}), companies tend to find ways to evade ad-blockers.
%
Due to space, we elaborate more on various countermeasures regarding readily available tools such as private browsers and ad-blockers in Appendix~\ref{sec:appendix-countermeasures}.
Alas, the question of how to preserve user privacy under behavioral tracking in the post-third-party cookie era, remains open.

%% file: sections/appendix_extended_results.tex
\section{Appendix: Extended Results}

\subsection{Categories adopting \FBP}
\label{sec:appendix-categories-adoption}

One question a careful reader might ask would be:
\textit{If a user, visits a website of category X (e.g. Shopping), what is the chance that the website will be utilizing Pixel?}

\begin{figure}[H]
\centering
    \begin{subfigure}[t]{0.45\textwidth}
        \vspace{-.2cm} \includegraphics[width=\columnwidth,height=5cm]{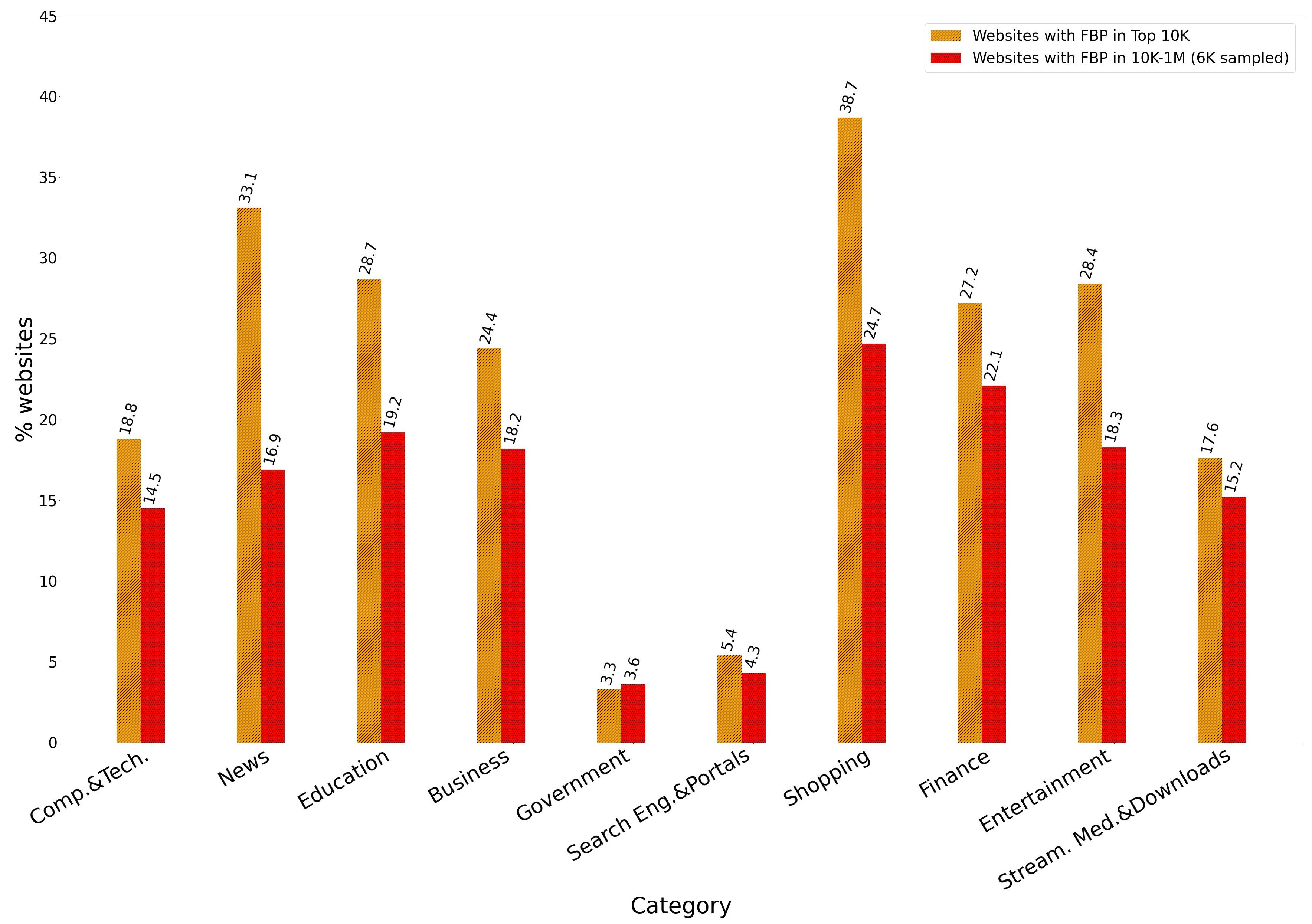}
        \vspace{-.6cm}
        \caption{\FBpixel adoption rate for the \TrancoFBP and \SampledFBP.}
        \Description{
            Category adoption rate for websites utilizing Facebook Pixel in the top 10K and 6K sampled lists. The adoption rate bars for the websites utilizing Pixel in the top 10K and the 6K sampled from the tranco list depict the population of websites utilizing Facebook Pixel with respect to the total population of websites of this category.
        }
        \label{fig:CategoryProbabilityPixel} 
    \end{subfigure}
    \begin{subfigure}[b]{0.45\textwidth}
       \includegraphics[width=\columnwidth]{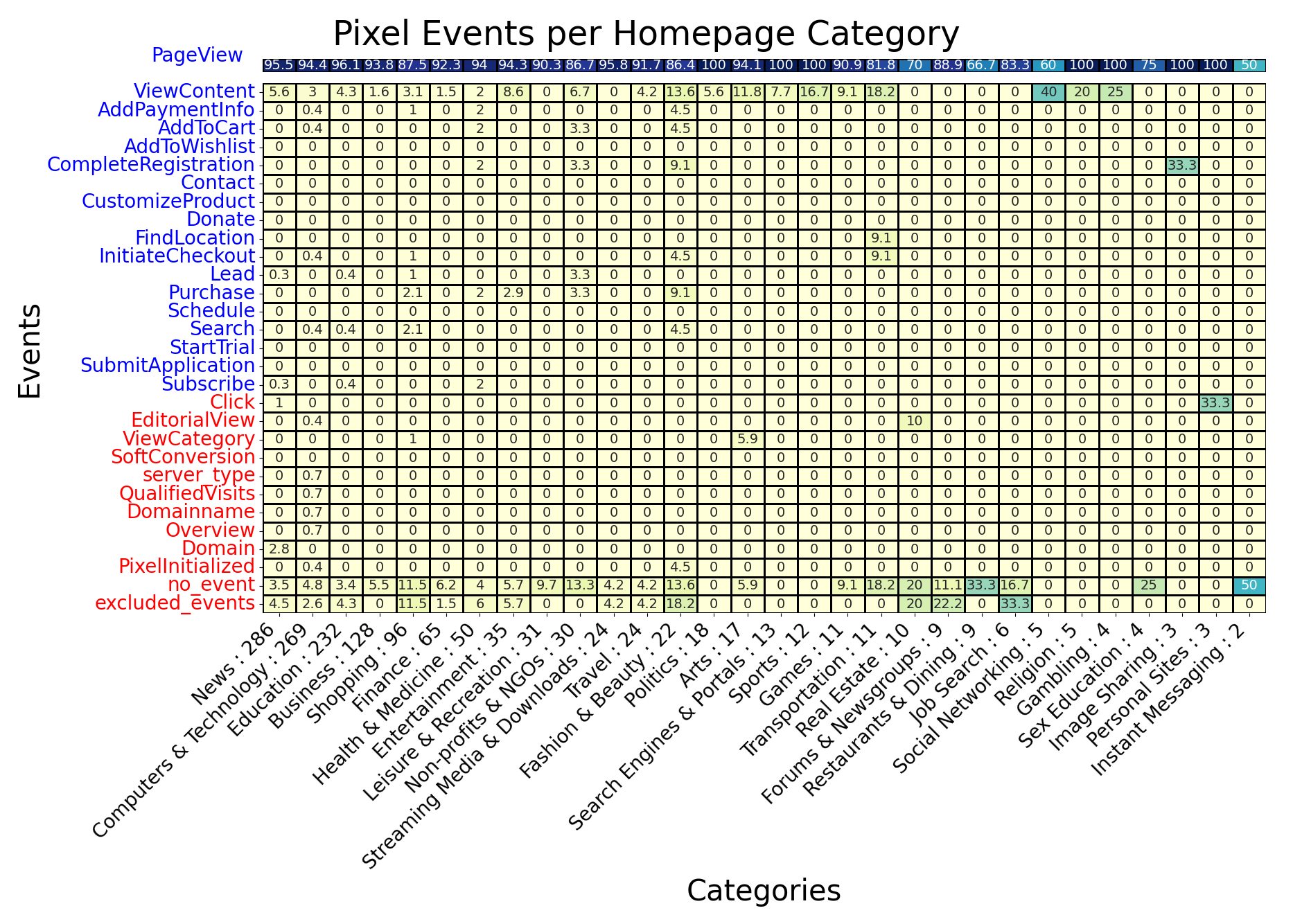}
        \vspace{-0.6cm}
        \caption{Events tracked by \FBpixel on homepages of \TopTranco websites, analyzed per category. Blue-colored events are the 18 standard from FB. Red-colored events are newly discovered and studied. Cells show portion of websites per category (x-axis) tracking said event (y-axis).}
        \label{fig:events-per-category-homepages}
        \Description{
            Heatmap depicting the portion of Standard and Custom events captured in the homepages of the websites utilizing Facebook Pixel across the tranco's top 10K. As with the sub-pages, portions are normalized with respect to websites' categories. 
        }
        \vspace{-4.5mm}
    \end{subfigure}
\end{figure}

To answer this question we need to quantify the adoption rate per category, for each examined category.
This differs from the results shown in Fig.~\ref{fig:GeneralWebVSPixelCategories} on the categorization of websites utilizing \FBpixel, and allows us to find what website categories are the most dominant across \FBpixel utilizing websites.
For example, according to Fig.~\ref{fig:CategoryProbabilityPixel}, 38.7\% of ``Shopping'' websites utilizes \FBpixel, but only 7.1\% of the total \TrancoFBP websites are of the ``Shopping'' category.
This means that, although \FBpixel is more widespread across ``Shopping'' websites, this category has a small population across websites found utilizing \FBpixel on the \TopTranco.
(Note: Fig.~\ref{fig:CategoryProbabilityPixel} results per category are sorted based on the order of Fig.~\ref{fig:GeneralWebVSPixelCategories} for easy comparison).

As the results in Fig.~\ref{fig:CategoryProbabilityPixel} indicate, the most predominant category that utilizes \FBpixel is ``Shopping'' websites in both sets, \TopTranco and \SampledTranco, with adoption rates 38.7\% and 24.7\% respectively.
This category is followed by News (33.1\%) and Education (28.7\%) for the websites utilizing Pixel from \TopTranco.For the websites utilizing Pixel from the \SampledTranco, the next two most predominant categories are Shopping (24.7\%) and Finance (22.1\%).

\subsection{Facebook Pixel Events \& User Data}
\label{sec:appendix-events}
In Fig.~\ref{fig:events-per-category-homepages}, we see that although websites can track various events on their pages, the most common is the 'PageView'.
As mentioned, \FBpixel allows a website to define the events to be tracked, by specifying each event in the fbq().
For example, the submission of a registration form can be tracked by 
\textit{fbq('track',CompleteRegistration)}.
Standard Events (as defined by \FB~\cite{pixel-events-documentation}) are listed below:
\vspace{-.4cm}
\begin{multicols}{3}
\begin{itemize}
    \item \footnotesize{PageView}
    \item \footnotesize{AddToCart}
    \item \footnotesize{Contact}
    \item \footnotesize{Donate}
    \item \footnotesize{Lead}
    \item \footnotesize{Purchase}
    \item \footnotesize{Schedule}
    \item \footnotesize{Search}
    \item \footnotesize{StartTrial}
    \item \footnotesize{Subscribe}
    \item \footnotesize{ViewContent}
\end{itemize}
\end{multicols}
\vspace{-.8cm}
\setlength{\columnsep}{-64pt}
\begin{multicols}{2}
\begin{itemize}
    \item \footnotesize{AddToWishlist}
    \item \footnotesize{AddPaymentInfo}
    \item \footnotesize{FindLocation}
    \item \footnotesize{CustomizeProduct}
    \item \footnotesize{CompleteRegistration}
    \item \footnotesize{InitiateCheckout}
    \item \footnotesize{SubmitApplication}
\end{itemize}
\end{multicols}
\vspace{-.4cm}

Alongside to the \fbp and \FBCLID (\FBC) values, as previously mentioned, the \FBpixel shares a variety of additional information regarding the user that caused the event, in order to enhance \FB attribution of activity to a real user.
Table~\ref{tab:user-data-fb} lists some of the extra user data (complete list can be found at~\cite{customer-information-parameters}) that can be reported by websites to \FB for better user matching.

\begin{table}[H]
\vspace{-3.5mm}
\caption{Sample of User data reported by websites to \FB for better user matching.
HR: Hashing Required; HRC: Hashing Recommended; DNH: Do Not Hash.}
\tiny{
\begin{tabular}{p{2.2cm}p{4cm}p{0.8cm}}
\toprule
Query parameter name & Description  & Hashing Guide\\
\midrule
em & Email & (HR)\\
ph & Phone Number & (HR)\\
fn & First Name & (HR)\\
ln & Last Name & (HR)\\
db & Date of birth & (HR)\\
ge & Gender & (HR)\\
ct & City & (HR)\\
st & State & (HR)\\
zp & Zip Code & (HR)\\
country & Country & (HR)\\
\midrule
fbp & The \FBP cookie value & (DNH)\\
fbc & The \FBC cookie value & (DNH)\\
\bottomrule
\end{tabular}
\label{tab:user-data-fb}
\Description{
    User Data: the available identifiers Facebook provides, that a website can define to be reported through Pixel, enhancing the matching and process. Each identifier has a query parameter name that will be included in the Pixel's request URL, a description and hashing requirement description (shared plain-text or hashed).
}
}
\end{table}

\begin{figure}[H]
\vspace{-0.5cm}
\centering
\begin{subfigure}[t]{0.42\textwidth}
   \includegraphics[width=\columnwidth]{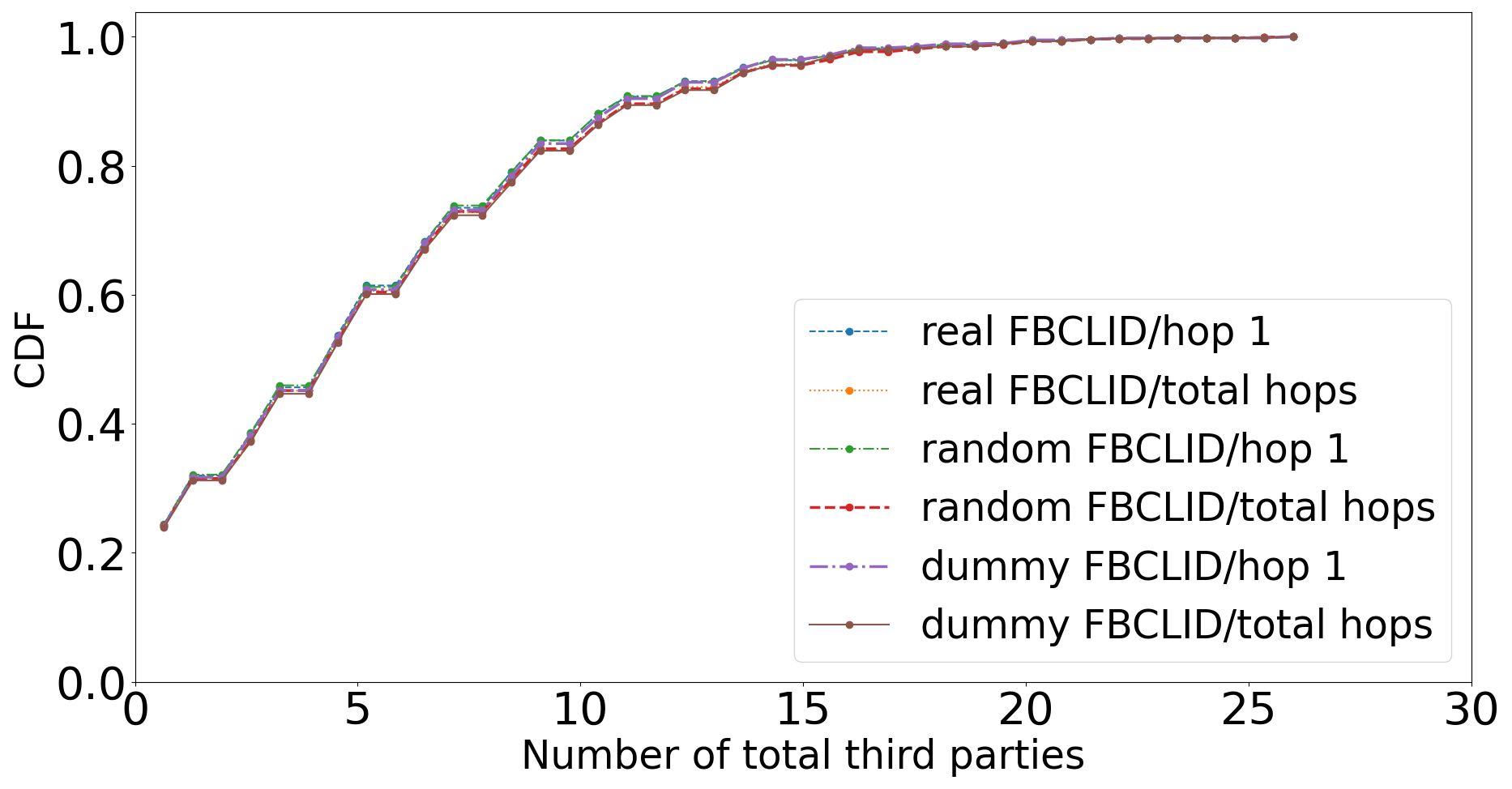}\vspace{-0.2cm}
    \caption{CDF of number of unique third parties informed of FB-related IDs, per hop \SampledFBP.}
    \label{fig:FbclidCDFSampled}
    \Description{
        Cumulative distribution function (CDF) of unique third parties per hop, receiving each FBCLID value (real, random, or dummy) for the 6K sampled websites utilizing Pixel. By unique third party, we refer to a third party domain that was not encountered in an earlier hop. Similar with the websites utilizing Pixel in the top 10K of the tranco list, the plot indicates that neither FBPixel nor websites handle the FBCLID tag differently under variations of the value.
    }
\end{subfigure}
\begin{subfigure}[b]{0.42\textwidth}
   \includegraphics[width=\columnwidth]{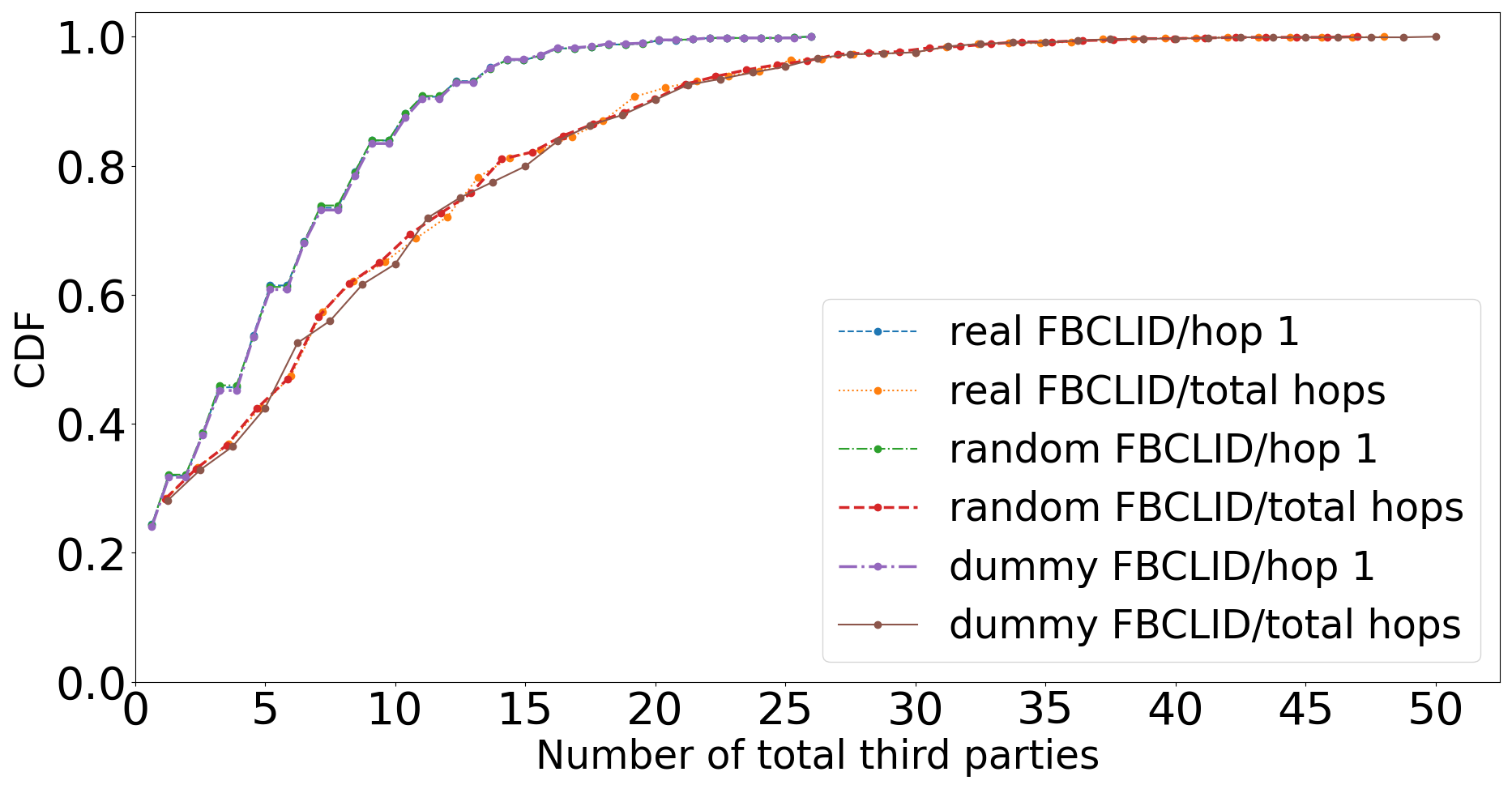}
    \caption{CDF of number of third parties informed of FB-related IDs, per hop, disregarding if they appeared in earlier hop in \SampledFBP.}
    \label{fig:FbclidDefaultDistributionPlotSampled}
    \Description{
       Cumulative distribution function (CDF) of third parties receiving each FBCLID value (real, random, or dummy), for the websites utilizing Pixel in the 6K sampled websites of the Tranco list. This time we plot the third parties per hop disregarding if they appeared on an earlier hop. 
    }
    \vspace{-3.5mm}
\end{subfigure}
\end{figure}

\subsection{\FBCLID Sharing With Third Parties}
\label{sec:appendix-FBCLIDsharing-third-party-distribution}

While visiting and monitoring the traffic on \SampledFBP websites coming from the top 10K-1M, the third-party invocations of unique and total numbers are found in Figures~\ref{fig:FbclidCDFSampled} and Figure~\ref{fig:FbclidDefaultDistributionPlotSampled}, respectively.

\begin{figure}[t]
    \centering
    \includegraphics[width=0.8\columnwidth]{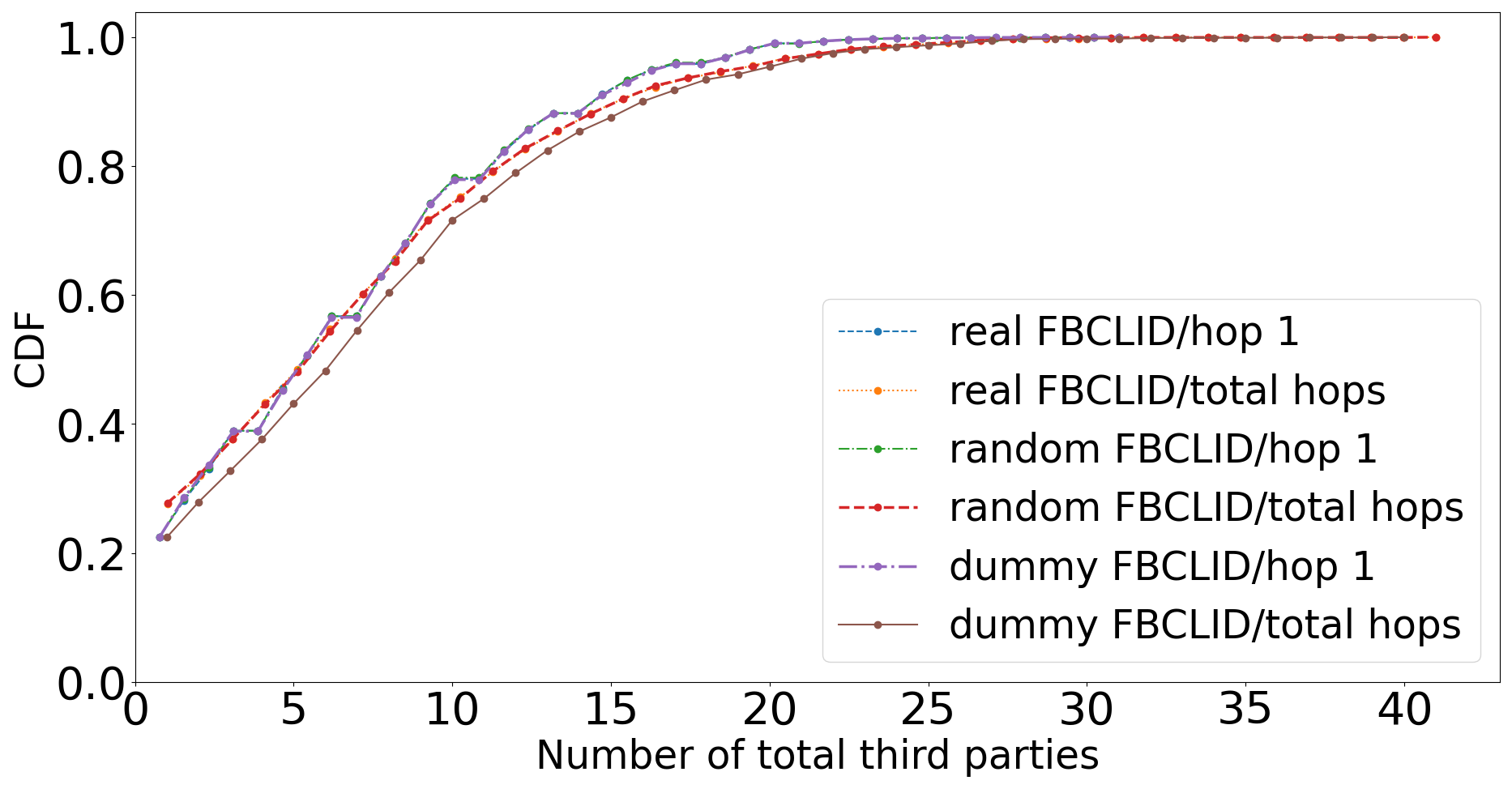}
    \vspace{-2mm}
    \caption{CDF of number of third parties informed of \FB-related IDs, per hop, disregarding if they appeared in earlier hop (\TrancoFBP).}\vspace{-0.2cm}
    \label{fig:FbclidDefaultDistributionPlot}
    \Description{
        Cumulative distribution of function (CDF) third parties receiving receiving each FBCLID value (real, random, dummy) for the websites utilizing Pixel in the top 10K of the Tranco list. This time we plot the third parties per hop disregarding if they appeared in an earlier hop. 
    }
\end{figure}

In Figure~\ref{fig:FbclidDefaultDistributionPlot}, we see the distribution of third parties per hop, disregarding if these third parties appeared again in earlier hop (excluding \FB-related domains), for the 3 cases, and the two lists.``First hop'' and ``total hops'' have similar meanings as described in \ref{sec:fbp-fbclickid-third-parties}. Looking into the Figure~\ref{fig:FbclidDefaultDistributionPlot}, the main difference from the plot in Figure \ref{fig:FbclidCDFs} is the lines for ``total'' hops. In fact, we find that the median website passes \FBCLIDs to 6.7 (7.9) and up to 41 (50) third parties within the 2nd hop. Therefore, and in contrast to unique third parties, lower ranked websites pass \FBCLIDs to more entities within 2 hops.

\begin{figure}[t]
    \includegraphics[width=0.9\columnwidth]{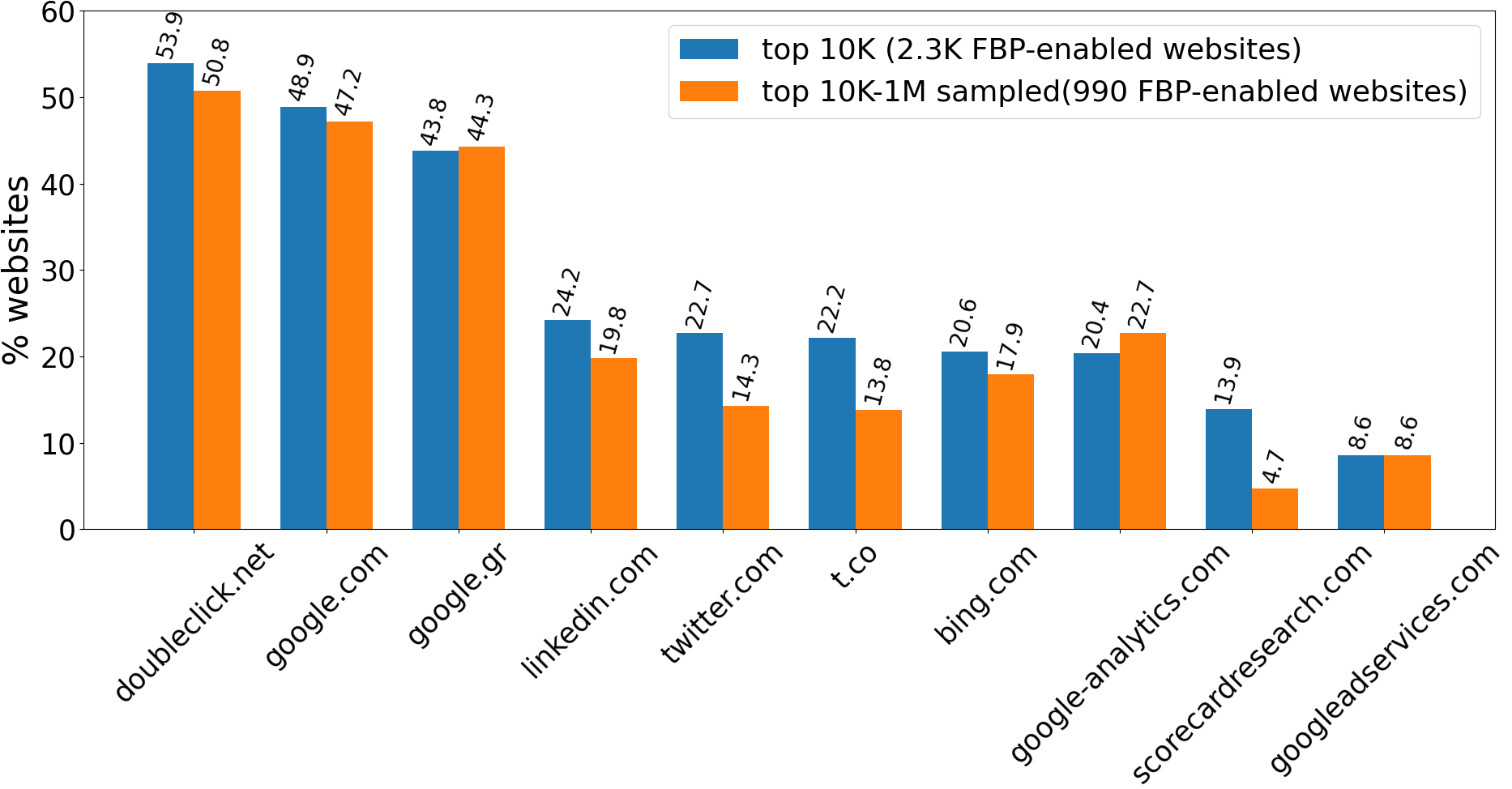}
    \caption{Top 10 third 
    parties from each list of \FBpixel-enabled websites (\TrancoFBP and \SampledFBP).}
    \label{fig:top-third-parties}
    \Description{
     Top 10 third party domains that receive Facebook Click IDs for each of the two lists of FBPixel-utilizing websites (i.e., websites utilizing Pixel in the top 10K and the 6K sampled from the Tranco list). In contrast with the entities Figure, here we plot the domains with the three most predominant being doubleclick.net, google.com and google.gr. 
    }
    \vspace{-3.5mm}
\end{figure}

Figure~\ref{fig:top-third-parties} shows the top 10 third parties informed with \FBCLIDs, ranked based on the portion of websites emitting these IDs from each list.
As no surprise, \texttt{doubleclick.net} and other top trackers are informed of a user's \FB activity from many websites.

\subsection{Countermeasures}
\label{sec:appendix-countermeasures}
Privacy by design browsers such as Brave, have utilized mechanisms to block those click ids and prevent query parameter tracking.
Specifically, Brave holds a list of known tracking parameters including the \FBCLID.
When a user visits a website with a tracking parameter, Brave removes the value from the URL before the browser makes the request.
So even if the website the user lands on utilizes \FBpixel, it will not be able to create a \FBC cookie and report it alongside with the \fbp.
One might assume that this would stop any potential matching of browsing activity (under the \fbp) to a real \FB profile.
But, as we mentioned, websites can declare other identifiers to be shared with \FBpixel such as the email, first \& last name, etc.
Although those identifiers are shared in a hashed form, they could still enable \FB to perform the matching process.

Furthermore, sharing of other unique identifiers, such as the SHA-256 hash of the user's email can be used to bypass mechanisms such as private/incognito browsing.
\FBpixel performs its event tracking even when the user browses in private mode (incognito).
By doing so, cookies and website data are recorded while the user is browsing in incognito, but deleted when they exit the browser's private mode.
This means that the persistent \fbp cookie will be created with a new value at each new visitation of a website and will be deleted after closing the browser.
So for each browsing duration, every action will be collected under a different \fbp (a different pseudonym). Someone could assume that this effort and process would break the ability of \FBP and \FB tracking and storing the history of activities of the same user under a unique pseudonym.
But again, the website can decide to share other \textbf{persistent} and \textbf{unique} identifiers such as the user's hashed email through \FBpixel, thus still enabling the correlation of multiple \fbp values with one (still) \textbf{anonymous} to \FB user.

Ad-blockers such Adblock Plus, UBlock, Disconnect, etc., can partially block the use of \FBpixel.
Such tools contain a list of filters which determine what third-party requests (e.g., resources such as third-party JavaScript) will be blocked. 
Adblock Plus, for example, blocks the request that the base \FBpixel code makes to fetch the \textit{fbevents} library.
Without this library, the \FBpixel cannot create the first-party cookies to either track or report events.
Although this action may restrain \FBpixel from performing the tracking correctly, it can still be bypassed.
Websites can utilize a variety of methods such as request redirection through proxy servers or similar techniques explained in~\cite{evading-adblockers}.
Moreover, the websites can utilize banners that require for the user to disable their ad-blocking extensions in order to continue browsing the website.

All these aforementioned efforts need to be investigated in the future in order to quantify how effective they can be, and if they are not, what other methods can be proposed by the security and privacy community.
Indeed, it seems that even though capturing browsing activity per site and user is becoming more expensive and difficult than in the past, large advertisers have both the capacity and resources to continue doing so.

\input{sections/13_ethics}

%% file: sections/13_ethics.tex
\subsection{Ethical Considerations}
\label{sec:appendix-ethics}



The execution of this work has followed the principles and guidelines of how to perform ethical information research and the use of shared measurement data~\cite{dittrich2012menloreport,rivers2014ethicalresearchstandards}.
In particular, this study paid attention to the following dimensions.
We keep our crawling to a minimum to ensure that we do not slow down or deteriorate the performance of any web service in any way, by visiting the websites a few times spread over a long period of time.
We do not interact with any component in the website visited, and only passively observe network traffic and cookies stored due to the visit.In addition to this, our crawler has been implemented to wait for both the website to fully load and an extra period of time before visiting another website.
Consequently, we emulate the behavior of a normal user that stumbled upon a website, and arrives with potentially some cookies and or URL parameters in their browser.
Therefore, we make a concerted effort not to perform any type of DoS attack on the visited websites. Also, and in accordance with GDPR and ePrivacy regulations, we did not engage in collection of data from real users.